\documentclass[journal]{IEEEtran}
\usepackage{amsfonts}
\usepackage{amssymb}
\usepackage{mathrsfs}
\usepackage{amsmath}
\usepackage{bm}
\usepackage{algorithmic}
\usepackage{algorithm}
\usepackage{enumitem}
\usepackage{graphicx}
\usepackage{subfigure}
\usepackage{float}
\usepackage{multirow} 
\usepackage{cite}
\usepackage[justification=centering]{caption}

\ifCLASSINFOpdf
\else
\fi

\begin{document}
\title{The PHD/CPHD filter for Multiple Extended Target Tracking with Trajectory Set Theory and Explicit Shape Estimation}
\author{Yuanhao Cheng,
	Yunhe Cao,~\IEEEmembership{Member,~IEEE,}
	Tat-Soon Yeo,~\IEEEmembership{Life~Fellow,~IEEE,}
	Jie Fu, Wei Zhang, Mengmeng Han
	\thanks{Yuanhao Cheng, Yunhe Cao, Fu Jie, and Wei Zhang are with the National Laboratory of Radar Signal Processing, Xidian University, Xi'an 710071, China.}
	\thanks{Tat-Soon Yeo is with the Department of Electrical and Computer Engineering, National University of Singapore, 119077, Singapore.}
        \thanks{Mengmeng Han is with the Beijing Electro-Mechanical Engineering Institute, Beijing 119077, China.}	
}

\markboth{Journal of \LaTeX\ Class Files,~Vol.~14, No.~8, August~2015}%
{Shell \MakeLowercase{\textit{et al.}}: Bare Demo of IEEEtran.cls for IEEE Journals}

\maketitle

\begin{abstract}
In this paper, we propose two methods for tracking multiple extended targets or unresolved group targets with elliptical extent shapes. These two methods are deduced from the Probability Hypothesis Density (PHD) filter and the Cardinality-PHD (CPHD) filter, respectively. In these two methods, Trajectory Set Theory (TST) is combined to establish the target trajectory estimates. Moreover, by employing a decoupled shape parameter model, the proposed methods can explicitly provide the shape estimation of the target, such as the orientation of the ellipse extension and the length of its two semi-axes. We derived the closed-form Bayesian recursions of these two methods with stable trajectory generation, referred to as the TPHD-E filter and the TCPHD-E filter. In addition, Gaussian mixture implementations of our methods with explicit extent update are provided, which are further named as the GM-TPHD-E filter and the GM-TCPHD-E filter. We illustrate the abilities of these methods through simulations and experiments with real data. These experiments demonstrate that the two proposed methods have advantages over existing filters in the accuracy of shape estimation, as well as in the completeness and correctness of target trajectory generation.
\end{abstract}


\IEEEpeerreviewmaketitle

\section{Introduction}
\IEEEPARstart{W}{ith} the development of sensor technology, Multiple Extended Target Tracking (METT) or Multiple Unresolvable Group Target Tracking (MUGTT) have become hot topics in the field of multi-target tracking \cite{2022Tutorial}. Related technologies are widely used in traffic monitoring \cite{2018ScheelTITS,2022LiuTIV}, unmanned technique \cite{2015KunzIV,2021MeyerTSP}, image vision \cite{2024XiongTIV}, biology \cite{2008SmalTMI}, and automation \cite{2017ChenICCAIS,2022HanumegowdaEuRAD}.

Unlike traditional multi-target tracking that treats a single target as a point, the METT (or MUGTT) also focuses on providing extent/contour estimations for multiple targets when tracking their kinematic states. The extent of the target is defined as a specific area where the target generates multiple measurements \cite{Granstrm2016ExtendedOT}. In addition to considering the extent of the target as lines \cite{2013GranströmFUSION}, splines \cite{2022DahlénFUSION}, triangles\cite{2025LiTAES}, rectangles \cite{2024CaoTITS}, or other complex shapes\cite{2021AlqaderiFusion,2015WahlströmTSP,2022AlqaderiTAES}, the most common approach currently is to consider the extended shape of the target as an ellipse. It has simplicity and can be applied to most real-life scenarios.

For targets with elliptical extents, the most frequently used shape modeling method is based on the Random Matrix Model (RMM) \cite{2008KochTAES}. It describes the elliptical shape as an unknown Semi Positive Definite Matrix (SPDM) and uses the inverse Wishart distribution to statistically analyze this SPDM. In RMM based target models, the overall state of the target is composed of Gaussian kinematic state and the inverse Wishart distributed extent matrix, presenting a Gaussian Inverse Wishart(GIW) distribution as a whole. This distribution has the property of conjugate prior, so it can be easily embedded into Bayesian filtering frameworks. Currently, combining the GIW distribution with multi-target tracking methods such as the Joint Probabilistic Data Association(JPDA) \cite{2015Schuster} or the Probabilistic Multi-Hypothesis Tracking (PMHT) \cite{2012Wieneke} is a feasible approach to achieve METT with a small number of targets. In tracking scenarios with numerous targets, a more representative implementation scheme is to combine the RMM with the Random Finite Set (RFS)-based methods such as the GIW-Probability Hypothesis Density (PHD) filter \cite{2012GranströmTSP}, the GIW-Cardinality-PHD (CPHD) filter \cite{2013LundquistJSTSP}, or other more advanced RFS-based filters \cite{2019GranströmTAES,2022XiaTAES}. Due to the efficiency of the RFS-based methods in multi-target tracking, this type of method has been widely applied in practice \cite{2015GranströmTGRS,2023LiuTITS}.

However, both of these methods that combine RMM and random finite sets have two main drawbacks. On the one hand, the RMM couples the shape parameters of a target in the extended matrix, and thus not able to explicitly provide the extent information of the ellipse, such as the axis length and orientation. In practical applications, such as obstacle avoidance of unmanned vehicles or cooperative flight of aircraft groups, shape information is an essential requirement. Moreover, RMM describes the changes in the extended shape based on a prior forgetting factor, which means that the accuracy of RMM in shape estimation is heavily dependent on manually specified initial values. On the other hand, the PHD/CPHD filter or other non-labeled RFS filters \cite{2012MallickBOOK} cannot directly provide the tracks in a mathematically rigorous way, which means that the methods proposed in \cite{2012GranströmTSP,2013LundquistJSTSP,2019GranströmTAES,2022XiaTAES,2015GranströmTGRS,2023LiuTITS} are not strictly trackers. 

Over the years, the above-mentioned methods have improved in two aspects: shape estimation models and trajectory building methods. In terms of trajectory building, many methods have been proposed to form target trajectories by adding labels, such as using tags for data association and track management \cite{2009PantaTAES}, or developing a series of label-based filters \cite{2016BeardTSP,2018DaniyanTSP}. These methods can handle general scenarios but do not work well in challenging situations, e.g., when targets move in the same direction and in close proximity. This can lead to problems such as ``trajectory switching'' \cite{2018GranströmFUSION}. The optimal trajectory construction methods are based on the Trajectory Set Theory (TST) \cite{2020GranströmFUSION}, which can generate complete and stable target trajectories within an acceptable time. Moreover, TST is highly applicable to RFS-based filters. The most representative method is proposed in \cite{2021SjudinFUSION}, which combines TST with the GIW-PHD filter to establish target trajectories. Other studies, such as \cite{2025WeiTAES}, consider coexisting point and extended target tracking based on \cite{2021SjudinFUSION}, and \cite{2023XiaTAES,2025GranströmTAES} combine TST with more advanced filters. 

However, these methods have not addressed the inherent drawbacks of RMM in shape estimation. In this regard, many improvements for the RMM model have been proposed, such as Granstr$\ddot{\mathrm{o}}$m's model \cite{2014GranströmTAES2} that improves shape prediction, Lan's model \cite{2016LanTAES} utilizes a shape evolution matrix, and Hoher's model \cite{2022HoherTSP} introduces a virtual measurement model. A series of shape state decoupling estimation models have also been developed, such as the Yang’s model \cite{2019YangTSP} based on the multiplicative noise term and the Tuncer’s model \cite{2021TuncerTSP} based on the variational inference. The Yang’s model has become a representative method due to its stability and lightweight, and its combination with JPDA can also be further used as a METT scheme \cite{2018Yang_LJPDAWorkshop,2020Yang_JIPDAFUSION}. However, both methods \cite{2018Yang_LJPDAWorkshop,2020Yang_JIPDAFUSION} have shortcomings in handling a large number of targets. \cite{2019YangTII} proposes to combine Yang’s model with the PHD filter to achieve METT, and uses network flow labeling to establish target trajectories. This seems to solve the two problems mentioned above, but the dynamic programming algorithm \cite{2011PirsiavashCVPR} is required to solve the network flow problem, which will consume a lot of computational resources. In addition, the network flow model requires application-specific customization and artificially specified parameters will greatly affect the generated track, so this method is not applicable in practice. 

In a word, there is currently no suitable method to solve the above two problems at the same time. In this work, we propose two METT schemes based on the PHD filter and the CPHD filter, respectively, with a decoupled parameter model to provide explicit shape estimation. The TST is employed and embedded into our methods, to further generate the target's trajectory. These two methods for METT are called the TPHD-E filter and the TCPHD-E filter. The main contributions of this paper are as follows.
\begin{itemize}
    \item In this paper, we propose two METT methods based on the PHD filter and the CPHD filter, respectively.
    \item The decoupled parameter model is introduced into our methods to provide an explicit extent information of the target, such as orientation, size. 
    \item  For the defects of standard PHD/CPHD filters in target trajectory generation, we developed our methods with the TST \cite{2019GarcíaTSP} and derived the closed form of recursion. The enhanced methods obtained can gracefully establish the target trajectory and serve as strictly defined trackers for extended targets, which are called the TPHD-E filter and the TCPHD-E filter.
    \item We also propose Gaussian mixture implementations of our TPHD-E/TCPHD-E filters with an explicit extent update to provide accurate shape estimation, which follow the spirit of the Gaussian mixture PHD/CPHD filters for extended targets \cite{2012GranströmTSP,2013LundquistJSTSP}. The probability hypothesis density in the obtained Gaussian Mixture TPHD-E (GM-TPHD-E) and TCPHD-E (GM-TCPHD-E) filters are composed of multiple trajectory Gaussian components.
    \item For the proposed methods, we further formulate a new pruning scheme that considers the kinematic and shape parameters of the target as decoupled parts and performs pruning judgments separately.
\end{itemize}

The remainder of the paper is organized as follows. Section \ref{Sec3} presents some background material on the problem definition, the TST and the Bayesian filtering framework. In Section \ref{Sec4}, we give the recursion of the TPHD-E filter and the TCPHD-E filter based on the TST for METT, and their implementations of the Gaussian mixture with an explicit extent update are provided in Section \ref{Sec 5}. Comparative experiments are shown in Section \ref{Sec6}. Finally, we draw some conclusions in Section \ref{Sec7}.‌

It should be noted that we did not specifically distinguish the concepts of ``the extended target'' and ``the unresolvable group target''. Although these two types of targets are slightly different in the coupling relationship between the kinematic state and the shape state, the methods proposed in this work are applicable to these two types of target. Therefore, in this context, we use the term `target' equivalent to refer to these two types of targets. For simplicity, the term `METT' also implies the concept of `MUGTT'.

\begin{table}[h]
\renewcommand{\arraystretch}{1.5}
\caption{NOTATIONS}
\label{tableNotations}
\vspace{-5mm}
\centering
\begin{tabular}{p{0.43\textwidth}}	
\hrulefill
\setlist[itemize]{leftmargin=*}
\begin{itemize}
\setlength{\itemsep}{0pt}
\setlength{\parsep}{0pt}
\setlength{\parskip}{0pt}
\item Real number field $\mathbb{R}$; Natural number field $\mathbb{N}$ 
\item Set of real horizontal vector of length n is represented with $\mathbb{R}^n$.
\item Set of real matrices of size $m\times n$ is represented with $\mathbb{R}^{m\times n}$.
\item $\mathcal{N}\left(\mathbf{x};\boldsymbol{\mu},\mathrm{\Sigma}\right)$ represents the multivariate Gaussian distribution with mean vector $\boldsymbol{\mu}\in\mathbb{R}^{n_\mathbf{x}}$ and covariance matrix $\mathrm{\Sigma}\in\mathbb{S}_{++}^{n_\mathbf{x}}$.

\item $\mathbf{I}_d$ indicates the $d$-order identity matrix, and $\mathbf{O}_d$ indicates the $d$-order zero matrix.
\item $\mathbf{1}_{a,b}$ denotes the one matrix with size $a\times b$, and $\mathbf{0}_{a,b}$ denotes the zero matrix with size $a\times b$.

\item $1_Y\left( X \right)$ is the generalized indicator function, and satisfies:
\begin{equation}
\nonumber
1_Y\left( X \right) \triangleq\begin{cases}
	1,  \mathrm{if} \ X\subseteq Y\\
	0,  \mathrm{otherwise}\\
\end{cases}
\end{equation}
\item  $\delta _{X,Y}$ is the Kronecker delta function, and satisfies:
\begin{equation}
\nonumber
\delta _{X,Y}\triangleq \begin{cases}
	1, \mathrm{if} \ X=Y\\
	0, \mathrm{otherwise}\\
\end{cases}
\end{equation}

\item $\binom{n}{m}$ represents calculating the combination $\mathsf{C}_{m}^{n}$:
\begin{equation}
\nonumber
    \mathsf{C}_{m}^{n}=\frac{n!}{\left[ m!\left( n-m \right) ! \right]}
\end{equation}

\item $a^\mathrm{T}$ means find the transpose of $a$.
\item $a\mathrm{!}$ means calculating the factorial of $a$.
\item $\mathrm{dim}\left( a \right)$ means taking the dimension of $a$.

\item diag[$a_1,a_2,\cdots,a_n$] returns the diagonal matrix whose diagonal elements are $a_1,a_2,\cdots,a_n$. 
\item blkdiag[$A_1,A_2,\cdots,A_n$] constructs block diagonal matrix with matrices $A_1,A_2,\cdots,A_n$.
\item $\mathcal{A} \times \mathcal{B}$ represents calculating the Cartesian product between set $\mathcal{A}$ and set $\mathcal{B}$.

\end{itemize} 
\hrulefill	
\end{tabular}
\end{table}

\section{BACKGROUND}
\label{Sec3}
In this section, we describe some background materials on using the Bayesian filtering framework and the TST for METT, such as the problem formula of METT, the introduction of TST, and a brief review of Bayesian recursion.

\subsection{Problem Formulation}
\label{3A}
Given a set of measurements $\mathcal{Z} _k=\left\{ \boldsymbol{z}_{k}^{1},\boldsymbol{z}_{k}^{2},...,\boldsymbol{z}_{k}^{m_k} \right\}$ at time step $k$, $m_k$ is the number of measurements. The METT aims to recursively estimate the states of multiple targets $\mathcal{E} _k=\left\{ \boldsymbol{\xi }_{k}^{\left( 1 \right)}\in \mathbb{R} ^{d_{\boldsymbol{\xi }}},\boldsymbol{\xi }_{k}^{\left( 2 \right)},...,\boldsymbol{\xi }_{k}^{\left( n_k \right)} \right\} \subseteq \mathscr{F} \left( \mathbb{R} ^{d_{\boldsymbol{\xi }}} \right)$, where $\mathscr{F} \left( \mathbb{R} ^{d_{\xi}} \right)$ represents all subsets of $\mathbb{R}^{d_{\xi}}$ and $n_k$ is the number of targets. As targets may appear or disappear, the number of targets in the surveillance area at each time step is uncertain.

The measurement set $\mathcal{Z}_k$ contains all the detection information captured by the sensor at time step $k$, including the measurements generated by the targets and the background clutter. Typically, the number of measurements generated by each target at the time step $k$ can be modeled using a Poisson distribution with parameter $\gamma \left( \xi _{k}^{\left( j \right)} \right), j=1,2,...,n_k$, and these measurements are assumed to be uniformly distributed on the surface of the corresponding target \cite{2005Kevin}. Similarly, clutter is assumed to be uniformly distributed in the surveillance area, and its quantity obeys the Poisson distribution with parameter $\lambda_k$. 

Considering a $d$-dimensional scenario, the augmented target state $\boldsymbol{\xi }_{k}^{\left( j \right)},j=1,2,...,n_k$ can be divided into two parts:

\begin{equation}
    \boldsymbol{\xi }_{k}^{\left( j \right)}=\left[ \left( \boldsymbol{r}_{k}^{\left( j \right)} \right) ^{\mathrm{T}},\left( \boldsymbol{s}_{k}^{\left( j \right)} \right) ^{\mathrm{T}} \right] ^{\mathrm{T}}
\end{equation}

\noindent where $\boldsymbol{r}_{k}^{\left( j \right)}$ refers to the kinematic state of the $j$-th target, which can include position, velocity, acceleration, and other information. Its dimension can be expressed as $d_{\boldsymbol{r}}=d_{o\boldsymbol{r}}\times d$, and $d_{o\boldsymbol{r}}$ is the dimension of the kinematic state in a one-dimensional physical space. $\boldsymbol{s}_{k}^{\left( j \right)}$ represents the shape state of the $j$-th target, including information on the extent such as size, contour, and orientation. Its dimension is $d_{\boldsymbol{s}}$, so the total dimension of the target state is $d_{\boldsymbol{\xi}}=d_{\boldsymbol{r}}+d_{\boldsymbol{s}}$. Moreover, the kinematic state and the shape state are treated independently.

In order to simplify the structured space model, the dynamic model of all targets can be considered to obey unified linear Gaussian, that is, for $j$-th target, there is:
\begin{equation}
\boldsymbol{r}_{k+1}^{\left( j\right)}=\boldsymbol{F}_{k+1|k}^{\boldsymbol{r}}\boldsymbol{r}_{k}^{\left( j \right)}+\boldsymbol{w}_{k+1}
\end{equation}
\begin{equation}
\boldsymbol{s}_{k+1}^{\left( j \right)}=\boldsymbol{F}_{k+1|k}^{\boldsymbol{s}}\boldsymbol{s}_{k}^{\left( j \right)}+\boldsymbol{v}_{k+1}
\end{equation}

\noindent where $\boldsymbol{F}_{k+1|k}^{\boldsymbol{r}}$ and $\boldsymbol{F}_{k+1|k}^{\boldsymbol{s}}$ represent the state transition matrices of the kinematic state and the shape state, respectively. $\boldsymbol{w}_{k+1}$ and $\boldsymbol{v}_{k+1}$ are Gaussian white noise, and their covariances are $\boldsymbol{Q}_{k+1|k}^{\boldsymbol{r}}$ and $\boldsymbol{Q}_{k+1|k}^{\boldsymbol{s}}$, respectively.

Each measurement of the target is deemed to originate from the measurement source and be subjected to Gaussian white noise interference. The measurement sources are assumed to be uniformly distributed on the target surface. For simplicity, the measurement process for each target is considered the same, so the measurement model of the $j$-th target is:
\begin{equation}
\boldsymbol{z}_{k}^{i}=\boldsymbol{H}_k\boldsymbol{r}_{k}^{\left( j \right)}+\boldsymbol{S}_{k}^{\left( j \right)}\boldsymbol{h}_{k}^{\left( j \right)}+\boldsymbol{e}_k
\end{equation}

\noindent where $\boldsymbol{S}_{k}^{\left( j \right)}$ is the shape matrix of the $j$-th target, which contains the shape size and orientation information. $\boldsymbol{h}_{k}^{\left( j \right)}$ is the multiplicative error term, and $\boldsymbol{e}_k$ represents additive Gaussian white noise with covariance $\boldsymbol{Q}_{k}^{\boldsymbol{e}}$.

\subsection{Sets of Trajectories}
The sequence of states of a target in time constitutes a trajectory \cite{2019GarcíaTSP}, and the length of the trajectory depends on its starting and ending time steps. Let $\boldsymbol{\chi }=\left( t,\boldsymbol{\xi }^{1:n} \right)$ define a trajectory with an initial time step $t$ and a length $n$, and $\boldsymbol{\xi }^{1:n}$ represents the sequence of target states over $n$ consecutive time steps $\boldsymbol{\xi }^{1:n}=\left( \boldsymbol{\xi }^1,\boldsymbol{\xi }^2,...,\boldsymbol{\xi }^n \right)$. 

All trajectories up to the time step $k$ form a space $\mathbb{T} _{\left( k \right)}=\uplus _{\left( t,n \right) \in \mathcal{I} \left( k \right)}\left\{ t \right\} \times \mathbb{R}^{nd_{\boldsymbol{\xi }}}$ with the set $\mathcal{I} \left( k \right) =\left\{ \left( t,n \right):0\leqslant t\leqslant k\,\,\mathrm{and} 1\leqslant n\leqslant k-t+1 \right\}$. And $\left( t,n \right)$ is the ``end identifier'' of a trajectory, the calculation `$\times $' here denotes the Cartesian product. Supposing there are  $n_k$ trajectories at time $k$, the set of trajectories is denoted as:
\begin{equation}
\mathcal{X} _k=\left\{ \boldsymbol{\chi }_1,\cdots ,\boldsymbol{\chi }_{n_k} \right\} \in \mathscr{F} \left( \mathbb{T} _{\left( k \right)} \right) 
\end{equation}
\noindent where $\mathscr{F} \left( \mathbb{T} _{\left( k \right)} \right)$ represents the set of all finite subsets of $\mathbb{T} _{\left( k \right)}$.

\subsection{Bayesian Filtering Recursion}
Given the posterior multi-trajectory density $\pi _{k-1}\left( \cdot \right)$ on the set of trajectories at the last time step $k-1$. With the measurement set $\mathcal{Z}_k$, the predicted multi-trajectory density $\omega _k\left( \cdot \right)$ and posterior multi-trajectory density $\pi _k\left( \cdot \right) $ at current time step $k$ can be obtained by the Chapman-Kolmogorov equation and the Bayes rule, respectively, i.e., 
\begin{equation}
\omega _k\left( \mathcal{X} _k \right) =\int{f\left( \mathcal{X} _k \middle| \mathcal{X} _{k-1} \right) \pi _{k-1}\left( \mathcal{X} _{k-1} \right)}\,\,d\mathcal{X} _{k-1}
\end{equation}

\begin{equation}
\pi _k\left( \mathcal{X} _k \right) =\frac{\varphi _k\left( \mathcal{Z} _k \middle| \mathcal{X} _k \right) \omega _k\left( \mathcal{X} _k \right)}{\int{\varphi _k\left( \mathcal{Z} _k \middle| \mathcal{X} _k \right) \omega _k\left( \mathcal{X} _k \right)}\,\,d\mathcal{X} _k}
\end{equation}

\noindent where $f\left( \cdot \middle| \cdot \right)$ is the transition density for the trajectories. Let $\tau _{\dot{k}}\left( \boldsymbol{\chi } \right) $ denotes the corresponding target state of trajectory $\boldsymbol{\chi }$ at a time step $k$, then the density of measurements of trajectories $\varphi _k\left( \mathcal{Z} _k \middle| \mathcal{X} _k \right) $ can be written as: 
\begin{equation}
\label{2.3.3}
\varphi _k\left( \mathcal{Z} _k \middle| \mathcal{X} _k \right) =\varphi _k\left( \mathcal{Z} _k \middle| \tau _k\left( \mathcal{X} \right) \right) 
\end{equation}
\noindent where $\tau _k\left( \mathcal{X} \right) $ denotes the corresponding multi-target state at the time $k$, and satisfies 
\begin{equation}
\tau _k\left( \mathcal{X} \right) =\bigcup_{\boldsymbol{\chi }\in \mathcal{X} _k}{\tau _k\left( \boldsymbol{\chi } \right)}
\end{equation}

Eq.(\ref{2.3.3}) points out that the measurement likelihood of the trajectory is only related to the state of the trajectory at the current time and the measurements. In addition, for convenience of reading, we provide a complete list of notations and corresponding explanations in Tab.\ref{tableNotations}.

\section{THE TST-BASED PHD/CPHD FILTER FOR EXTENDED TARGETS}
\label{Sec4}
In this section, we focus on introducing the two proposed methods, which are developed based on the PHD filter and the CPHD filter, respectively. They can be applied to METT and further estimate the trajectories of targets based on TST. The recursion of the resulting TPHD-E filter and TCPHD-E filter are presented in Section \ref{Sec4-A} and Section \ref{Sec4-B}, respectively. Among them, the recursive derivation of the TPHD-E filter is based on \cite{Granstrom2012ImplementationOT}, while the inference of the TCPHD-E filter is grounded in \cite{2011OrgunerFUSION}.

\subsection{The TST-Based PHD Filter for Extended Targets}
\label{Sec4-A}
Let $\mathscr{D} \left( \boldsymbol{\chi } \right) =\left( t,\boldsymbol{\xi }^{1:n} \right)$ represent the PHD of the multiple alive trajectory. The TST-Based PHD filter for Extended targets (TPHD-E filter) propagates the multi-trajectory density of a Poisson RFS through the filtering recursion. Given the posterior trajectory PHD $\mathscr{D} _{\pi _{k-1}}\left( \cdot \right)$ at time step $k-1$, and the birth PHD $\mathscr{D} _{\beta ^{\tau}}\left( \cdot \right)$. Then the predicted trajectory PHD $\mathscr{D} _{\omega _k}\left( \cdot \right) $ at time step $k$ is:
\begin{equation}
\label{3.1.1}
\mathscr{D} _{\omega _k}\left( \boldsymbol{\chi } \right) =\mathscr{D} _{\omega _k}^{\mathsf{S}}\left( \boldsymbol{\chi } \right) +\mathscr{D} _{\omega _k}^{\mathsf{B}}\left( \boldsymbol{\chi } \right) 
\end{equation}
\noindent where
\begin{equation}
\label{3.1.2}
\begin{split}
&\mathscr{D} _{\omega _k}^{\mathsf{S}}\left( t,\boldsymbol{\xi }^{1:n} \right) \\
&=p^S\left( \boldsymbol{\xi }^{n-1} \right) g\left( \boldsymbol{\xi }^n \middle| \boldsymbol{\xi }^{n-1} \right) \mathscr{D} _{\pi _{k-1}}\left( t,\boldsymbol{\xi }^{1:n-1} \right) 1_{\mathbb{N} _{k-1}}\left( t \right)  
\end{split}
\end{equation}
\begin{equation}
\label{3.1.3}
\mathscr{D} _{\omega _k}^{\mathsf{B}}\left( t,\boldsymbol{\xi }^{1:n} \right) =\mathscr{D} _{\beta ^{\tau}}\left( \boldsymbol{\xi }^1 \right) 1_{\left\{ k \right\}}\left( t \right) 
\end{equation}
if $t+n-1=k$ and zero, otherwise. 

Eq.(\ref{3.1.1}) shows the predicted trajectory PHD is the sum of the PHD of the trajectories $\mathscr{D} _{\omega _k}^{\mathrm{B}}\left( \boldsymbol{\chi } \right)$ born at the time step $k$ and the PHD of the surviving trajectories $\mathscr{D} _{\omega _k}^{\mathrm{S}}\left( \boldsymbol{\chi } \right) $. In Eq.(\ref{3.1.2}), $p^S\left( \cdot \right)$ is the survival probability of the trajectory. $g\left( \cdot \middle| \cdot \right) $ denotes the transition density. $1_Y\left( X \right)$ is the generalized indicator function. Note that in Eq.(\ref{3.1.3}), we use superscript `$\tau$' in the densities on the RFS of the targets.

Then given the prior trajectory PHD $\mathscr{D} _{\omega _k}\left( \cdot \right)$ at time step $k$, the posterior trajectory PHD $\mathscr{D} _{\pi _k}\left( \cdot \right)$ at time step $k$ can be expressed as:
\begin{equation}
\label{3.1.4}
\mathscr{D} _{\pi _k}\left( t,\boldsymbol{\xi }^{1:n} \right) =\mathscr{L} _{\mathcal{Z} _k}\left( \boldsymbol{\xi }^n \right) \mathscr{D} _{\omega _k}\left( t,\boldsymbol{\xi }^{1:n} \right) 
\end{equation}
if $t+n-1=k$ and zero, otherwise. Where $\mathscr{L} _{\mathcal{Z} _k}\left( \boldsymbol{\xi }^n \right)$ is the pseudo-likelihood function, which can be calculated as:
\begin{equation}
\label{3.1.5}
\mathscr{L} _{\mathcal{Z} _k}\left( \boldsymbol{\xi }^n \right) =\varphi \left( \oslash \middle| \boldsymbol{\xi }^n \right) +\sum_{\mathcal{P} \angle \mathcal{Z} _k}{w_{\mathcal{P}}\sum_{\mathcal{C} \in \mathcal{P}}{\frac{\varphi \left( \left( \mathcal{P} ,\mathcal{C} \right) \middle| \boldsymbol{\xi }^n \right)}{\varrho _{\left( \mathcal{P} ,\mathcal{C} \right)}+\varsigma _{\left( \mathcal{P} ,\mathcal{C} \right)}}}}
\end{equation}
with
\begin{equation}
\label{3.1.6}
\varsigma _{\left( \mathcal{P} ,\mathcal{C} \right)}=\int{\varphi \left( \left( \mathcal{P} ,\mathcal{C} \right) \middle| \boldsymbol{\xi }^n \right) \mathscr{D} _{\omega _{k}^{\tau}}\left( \boldsymbol{\xi }^n \right) d\boldsymbol{\xi }^n}
\end{equation}
\begin{equation}
\label{3.1.7}
\varrho _{\left( \mathcal{P} ,\mathcal{C} \right)}=\delta _{\left| \mathcal{P} ,\mathcal{C} \right|,1}\left[ \prod_{\boldsymbol{z}\in \left( \mathcal{P} ,\mathcal{C} \right)}{\lambda ^C\left( \boldsymbol{z} \right)} \right] , \left| \mathcal{P} ,\mathcal{C} \right|>0
\end{equation}
\begin{equation}
\label{3.1.8}
w_{\mathcal{P}}=\frac{\prod\nolimits_{\mathcal{C} \in \mathcal{P}}^{\,\,}{\left( \varrho _{\left( \mathcal{P} ,\mathcal{C} \right)}+\varsigma _{\left( \mathcal{P} ,\mathcal{C} \right)} \right)}}{\sum\nolimits_{\mathcal{P} \angle \mathcal{Z} _k}^{\,\,}{\prod\nolimits_{\mathcal{C} \in \mathcal{P}}^{\,\,}{\left( \varrho _{\left( \mathcal{P} ,\mathcal{C} \right)}+\varsigma _{\left( \mathcal{P} ,\mathcal{C} \right)} \right)}}}
\end{equation}

In Eq.(\ref{3.1.5}), $\mathcal{P} \angle \mathcal{Z} _k$ denotes the partitions of the measurement set $\mathcal{Z} _k$, which means that $\mathcal{Z} _k$ can be divided into multiple disjoint nonempty measurement cells with partition $\mathcal{P}$. Each measurement cell can be regarded as a subset of $\mathcal{Z} _k$, then $\mathcal{C} \in \mathcal{P} $ shows that the set $\mathcal{C}$ is one of the cells under the partition , and the subscript `$\left( \mathcal{P} ,\mathcal{C} \right) $' indicates the cell $\mathcal{C} $ under the partition $\mathcal{P}$. In Eq.(\ref{3.1.7}), $\delta _{X,Y}$ is the Kronecker delta function, $\left| \mathcal{P} ,\mathcal{C} \right|$ represents the number of measurements contained in cell $\left( \mathcal{P} ,\mathcal{C} \right)$, and $\lambda ^C\left( \cdot \right) $ is the intensity of the clutter. In addition, in Eq.(\ref{3.1.6}), $\mathscr{D} _{\omega _{k}^{\tau}}\left( \boldsymbol{\xi }^n \right)$ representing the PHD of the targets at time step $k$ of density $\omega _k\left( \cdot \right)$, which can be obtained by marginalizing the $\mathscr{D} _{\omega _k}\left( \cdot \right)$, and $\varphi \left( \cdot \middle| \boldsymbol{\xi }^n \right)$ indicates the target-generated measurement density, which can be written as:
\begin{equation}
\begin{split}
&\varphi \left( \mathcal{Z} \middle| \boldsymbol{\xi }^n \right)\\ &=\begin{cases}
1-\left( 1-e^{-\gamma \left( \boldsymbol{\xi }^n \right)} \right) p^{\mathrm{D}}\left( \boldsymbol{\xi }^n \right) \,\,                     \left| \mathcal{Z} \right|=0\\
p^{\mathrm{D}}\left( \boldsymbol{\xi }^n \right) \gamma ^{\left| \mathcal{Z} \right|}\left( \boldsymbol{\xi }^n \right) e^{-\gamma \left( \boldsymbol{\xi }^n \right)}\prod\nolimits_{\boldsymbol{z}\in \mathcal{Z}}^{\,\,}{\varphi _{\boldsymbol{z}}\left( \boldsymbol{\xi }^n \right)}\,\,     \left| \mathcal{Z} \right|>0\\  
\end{cases}
\end{split}
\end{equation}
\noindent where $\varphi _{\boldsymbol{z}}\left( \cdot \right)$ is the likelihood of a single target-generated measurement, $p^D\left( \cdot \right)$ is the detection probability of the trajectory.

\subsection{The TST-based CPHD Filter for Extended Targets}
\label{Sec4-B}
Unlike the TPHD-E filter, the TST-Based CPHD filter for Extended targets (TCPHD-E filter) propagates multi-trajectory density of an independent identically distributed RFS through the recursion Bayesian filtering. Specifically, the TCPHD-E filter has an additional cardinality distribution $\mathscr{P} _{k|k}\left( \cdot \right)$, which is the Probability Mass Function (PMF) of the cardinality of the trajectory set. And $\mathscr{P} _{k|k}\left( n \right)$ represents the possibility that the number of alive trajectories is $n$. Therefore, the TCPHD-E filter is more accurate in estimating the number of targets than the TPHD-E filter.

Assume that the set of new born trajectories at time $k$ has cardinality $\mathscr{P} _{k}^{\beta}\left( \cdot \right) =\mathscr{P} ^{\beta}\left( \tau _k\left( \cdot \right) \right) =\mathscr{P} _{\beta _{k}^{\tau}}\left( \cdot \right) $ and PHD:
\begin{equation}
\mathscr{D} _{k}^{\beta}\left( t,\boldsymbol{\xi }^{1:n} \right) =\begin{cases}
\mathscr{D} _{\beta ^{\tau}}\left( \boldsymbol{\xi }^1 \right) \,\,   t=k,i=1\\
0       \qquad      \mathrm{otherwise}\\
\end{cases}
\end{equation}

Based on \cite{2019GarcíaTSP}, the PHD of the predicted density is the same as Eq.(\ref{3.1.1}). And the cardinality distribution of the predicted density is:
\begin{equation}
\mathscr{P} _{\omega _k}\left( m \right) =\sum_{j=0}^m{\left[ \begin{array}{c}
	\mathscr{P} _{k}^{\beta}\left( m-j \right) \sum_{l=j}^{\infty}{\binom{l}{j}}\mathscr{P} _{\pi _{k-1}}\left( l \right)\\
	{\begin{array}{c}
	\times \frac{\left[ \int{\left( 1-p^{S}\left( \boldsymbol{\xi } \right) \right) \mathscr{D} _{\pi _{k-1}^{\tau}}\left( \boldsymbol{\xi } \right) d\boldsymbol{\xi }} \right] ^{l-j}}{\left[ \int{\mathscr{D} _{\pi _{k-1}^{\tau}}\left( \boldsymbol{\xi } \right) d\boldsymbol{\xi }} \right] ^l}\\
	\times \left[ \int{p^{S}\left( \boldsymbol{\xi } \right) \mathscr{D} _{\pi _{k-1}^{\tau}}\left( \boldsymbol{\xi } \right) d\boldsymbol{\xi }} \right] ^j\\
\end{array}}\\
\end{array} \right]}
\end{equation}
where $\mathscr{D} _{\pi _{k-1}^{\tau}}\left( \boldsymbol{\xi } \right)$ is the PHD of the targets at time step $k-1$ according to $\pi _{k-1}\left( \cdot \right) $, which can be calculated as Eq.(\ref{3.1.4}). $\binom{l}{j}$ represents calculating the combination $\mathsf{C}_{j}^{l}$.

The PHD of the posterior density at time step $k$ can be calculated by referring to \cite[Eq.(10)]{2013LundquistJSTSP}, but the difference is that the posterior PHD here contains information about previous states of the trajectories, which can be written as Eq.(\ref{3.2.3}). And the posterior cardinality distribution $\mathscr{P} _{k|k}\left( n \right) $ at time step $k$ can be calculated according to \cite[Eq.(11)]{2013LundquistJSTSP}, shown in Eq.(\ref{3.2.4}).
\begin{figure*}[!t]
\begin{equation}
\label{3.2.3}
\scriptsize
\mathscr{D} _{\pi _k}\left( t,\boldsymbol{\xi }^{1:n} \right) =\left\{ \begin{matrix}
	\left[ \kappa \left( 1-p^D\left( \boldsymbol{\xi }^n \right) +p^D\left( \boldsymbol{\xi }^n \right) \mathscr{G} _{\boldsymbol{z}}\left( 0 \middle| \boldsymbol{\xi }^n \right) \right) +p^{\mathrm{D}}\left( \boldsymbol{\xi }^n \right) \frac{\sum_{\mathcal{P} \angle \mathcal{Z} _k}{\sum_{\mathcal{C} \in \mathcal{P}}{\left( \nu _{\mathcal{P} ,\mathcal{C}}\mathscr{G} _{\boldsymbol{z}}^{\left( \left| \mathcal{P} ,\mathcal{C} \right| \right)}\left( 0 \middle| \boldsymbol{\xi }^n \right) \prod_{\boldsymbol{z}\in \left( \mathcal{P} ,\mathcal{C} \right)}{\frac{\varphi _{\boldsymbol{z}}\left( \boldsymbol{\xi }^n \right)}{\varphi _{\mathrm{FA}}\left( \boldsymbol{z} \right)}} \right)}}}{\sum_{\mathcal{P} \angle \mathcal{Z} _k}{\sum_{\mathcal{C} \in \mathcal{P}}{\vartheta _{\mathcal{P} ,\mathcal{C}}\epsilon _{\mathcal{P} ,\mathcal{C}}}}} \right] {\frac{\mathscr{D} _{\omega _k}\left( t,\boldsymbol{\xi }^{1:n} \right)}{\int{\mathscr{D} _{\omega _{k}^{\tau}}\left( \boldsymbol{\xi } \right) d\boldsymbol{\xi }}}}\,\,&		\left| \mathcal{Z} _k \right|>0\\
	\left( 1-p^D\left( \boldsymbol{\xi }^n \right) +p^D\left( \boldsymbol{\xi }^n \right) \mathscr{G} _{\boldsymbol{z}}\left( 0 \middle| \boldsymbol{\xi }^n \right) \right) \mathscr{D} _{\omega _k}\left( t,\boldsymbol{\xi }^{1:n} \right)&		\left| \mathcal{Z} _k \right|=0\\
\end{matrix} \right. 
\end{equation}
\end{figure*}

\begin{figure*}[!t]
\begin{equation}
\label{3.2.4}
\small
\mathscr{P} _{k|k}\left( n \right) =\left\{ \begin{matrix}
	\frac{\sum_{\mathcal{P} \angle \mathcal{Z} _k}{\sum_{\mathcal{C} \in \mathcal{P}}{\vartheta _{\mathcal{P} ,\mathcal{C}}\mathscr{G} _{k|k-1}^{\left( n \right)}\left( 0 \right) \left( \mathscr{G} _{\mathrm{FA}}\left( 0 \right) \frac{\varpi _{\mathcal{P} ,\mathcal{C}}}{\left| \mathcal{P} \right|}\frac{\upsilon ^{n-\left| \mathcal{P} \right|}}{\left( n-\left| \mathcal{P} \right| \right) !}\delta _{n\geqslant \left| \mathcal{P} \right|}+\mathscr{G} _{\mathrm{FA}}^{\left( \left| \mathcal{P} ,\mathcal{C} \right| \right)}\left( 0 \right) \frac{\upsilon ^{n-\left| \mathcal{P} \right|+1}}{\left( n-\left| \mathcal{P} \right|+1 \right) !}\delta _{n\geqslant \left| \mathcal{P} \right|-1} \right)}}}{\sum_{\mathcal{P} \angle \mathcal{Z} _k}{\sum_{\mathcal{C} \in \mathcal{P}}{\vartheta _{\mathcal{P} ,\mathcal{C}}\epsilon _{\mathcal{P} ,\mathcal{C}}}}}&		\left| \mathcal{Z} _k \right|\ne 0\\
	\frac{\upsilon ^n\mathscr{G} _{k|k-1}^{\left( n \right)}\left( 0 \right)}{\mathscr{G} _{k|k-1}\left( \upsilon \right)}&		\left| \mathcal{Z} _k \right|=0\\
\end{matrix} \right. 
\end{equation}
\hrulefill
\end{figure*}

In Eq.(\ref{3.2.3}) and Eq.(\ref{3.2.4}), $\mathscr{G} _{\mathrm{FA}}\left( \cdot \right)$, $\mathscr{G} _{\boldsymbol{z}}\left( \cdot \middle| \boldsymbol{\xi } \right) $ and $\mathscr{G} _{k|k-1}\left( \cdot \right)$ denote the Probability Generating Function (PGF) of the false alarms, the PGF of the measurements conditioned on the state, and the predicted PGF of the state, respectively. Their calculation can be found in \cite{2013LundquistJSTSP}, and the superscript `$\left( n \right) $' of them indicates the $n$-order derivation. The coefficients $\kappa$, $\upsilon$ and variables $\varpi _{\mathcal{P} ,\mathcal{C}}$, $\eta _{\mathcal{P} ,\mathcal{C}}$, $\vartheta _{\mathcal{P} ,\mathcal{C}}$, $\epsilon _{\mathcal{P} ,\mathcal{C}}$, $\nu _{\mathcal{P} ,\mathcal{C}}$ can be calculated as:
\begin{equation}
\label{3.2.5}
\kappa \triangleq \frac{\sum\nolimits_{\mathcal{P} \angle \mathcal{Z} _k}^{\,\,}{\sum\nolimits_{\mathcal{C} \in \mathcal{P}}^{\,\,}{\vartheta _{\mathcal{P} ,\mathcal{C}}\mu _{\mathcal{P} ,\mathcal{C}}}}}{\sum\nolimits_{\mathcal{P} \angle \mathcal{Z} _k}^{\,\,}{\sum\nolimits_{\mathcal{C} \in \mathcal{P}}^{\,\,}{\vartheta _{\mathcal{P} ,\mathcal{C}}\epsilon _{\mathcal{P} ,\mathcal{C}}}}}
\end{equation}
\begin{equation}
\label{3.2.6}
\upsilon \triangleq \frac{\int{\left( 1-p^D\left( \boldsymbol{\xi } \right) +p^D\left( \boldsymbol{\xi } \right) \mathscr{G} _{\boldsymbol{z}}\left( 0 \middle| \boldsymbol{\xi } \right) \right) \mathscr{D} _{\omega _{k}^{\tau}}\left( \boldsymbol{\xi } \right) d\boldsymbol{\xi }}}{\int{\mathscr{D} _{\omega _{k}^{\tau}}\left( \boldsymbol{\xi } \right) d\boldsymbol{\xi }}}
\end{equation}
\begin{equation}
\label{3.2.7}
\vartheta _{\mathcal{P} ,\mathcal{C}}\triangleq \prod_{\mathcal{C} ^{\prime}\in \left( \mathcal{P} -\mathcal{C} \right)}{\varpi _{\mathcal{P} ,\mathcal{C} ^{\prime}}}
\end{equation}
\begin{equation}
\label{3.2.8}
\varpi _{\mathcal{P} ,\mathcal{C}}\triangleq \frac{\int{p^{{D}}\left( \boldsymbol{\xi } \right) \mathscr{G} _{\boldsymbol{z}}^{\left( \left| \mathcal{P} ,\mathcal{C} \right| \right)}\left( 0 \middle| \boldsymbol{\xi } \right) \prod_{\boldsymbol{z}\prime\in \mathcal{C}}{\frac{\varphi _{\boldsymbol{z}\prime}\left( \boldsymbol{\xi } \right)}{\varphi _{\mathrm{FA}}\left( \boldsymbol{z}\prime \right)}}\mathscr{D} _{\omega _{k}^{\tau}}\left( \boldsymbol{\xi } \right) d\boldsymbol{\xi }}}{\int{\mathscr{D} _{\omega _{k}^{\tau}}\left( \boldsymbol{\xi } \right) d\boldsymbol{\xi }}}
\end{equation}
\begin{equation}
\label{3.2.9}
\epsilon _{\mathcal{P} ,\mathcal{C}}\triangleq\mathscr{G} _{\mathrm{FA}}\left( 0 \right) \mathscr{G} _{k|k-1}^{\left( \left| \mathcal{P} \right| \right)}\left( \upsilon \right) \frac{\varpi _{\mathcal{P} ,\mathcal{C}}}{\left| \mathcal{P} \right|}+\mathscr{G} _{\mathrm{FA}}^{\left| \mathcal{P} ,\mathcal{C} \right|}\left( 0 \right) \mathscr{G} _{k|k-1}^{\left( \left| \mathcal{P} \right|-1 \right)}\left( \upsilon \right) 
\end{equation}
\begin{equation}
\label{3.2.10}
\mu _{\mathcal{P} ,\mathcal{C}}\triangleq\mathscr{G} _{\mathrm{FA}}\left( 0 \right) \mathscr{G} _{k|k-1}^{\left( \left| \mathcal{P} \right|+1 \right)}\left( \upsilon \right) \frac{\varpi _{\mathcal{P} ,\mathcal{C}}}{\left| \mathcal{P} \right|}+\mathscr{G} _{\mathrm{FA}}^{\left| \mathcal{P} ,\mathcal{C} \right|}\left( 0 \right) \mathscr{G} _{k|k-1}^{\left( \left| \mathcal{P} \right| \right)}\left( \upsilon \right) 
\end{equation}
\begin{equation}
\label{3.2.11}
\nu _{\mathcal{P} ,\mathcal{C}}\triangleq \frac{\vartheta _{\mathcal{P} ,\mathcal{C}}}{\left| \mathcal{P} \right|}\mathscr{G} _{\mathrm{FA}}\left( 0 \right) \mathscr{G} _{k|k-1}^{\left( \left| \mathcal{P} \right| \right)}\left( \upsilon \right) +\frac{\sum\nolimits_{\mathcal{C} \prime\in \left( \mathcal{P} -\mathcal{C} \right)}^{\,\,}{\vartheta _{\mathcal{P} ,\mathcal{C} \prime}\epsilon _{\mathcal{P} ,\mathcal{C} \prime}}}{\varpi _{\mathcal{P} ,\mathcal{C}}}
\end{equation}

In Eq.(\ref{3.2.4}), `$\left| \mathcal{P} \right|$' denotes the number of measurement cells in the partition $\mathcal{P}$, and the notation `$\left( \mathcal{P} -\mathcal{C} \right) $' in Eq.(\ref{3.2.7}) and Eq.(\ref{3.2.11}) indicates all measurement cells except $\left( \mathcal{P} ,\mathcal{C} \right) $ in the partition $\mathcal{P}$.

\section{GAUSSIAN MIXTURE IMPLEMENTATIONS WITH EXPLICIT EXTENT UPDATE}
\label{Sec 5}
In this section, we present the Gaussian mixture implementations of the TPHD-E and TCPHD-E filters with an explicit extent update. And the trajectory Gaussian density can be denoted as:
\begin{equation}
\mathcal{N} \left( \boldsymbol{\chi };t_k,\dot{\boldsymbol{\xi}}_k,\dot{\boldsymbol{\varXi}}_k \right) =\left\{ \begin{matrix}
\mathcal{N} \left( \boldsymbol{\xi }^{1:n}; \dot{\boldsymbol{\xi}}_k,\dot{\boldsymbol{\varXi}}_k \right)&		t=t_k,n=n_k\\
	0&		\mathrm{otherwise}\\
\end{matrix} \right. 
\end{equation}

\noindent where $n_k={{\mathrm{dim}( \dot{\boldsymbol{\xi}}_k )}/{d_{\boldsymbol{\xi }}}}$, and $\left( t_k,n_k \right) $ represent the end indicators of the trajectory. $\dot{\boldsymbol{\xi}}_k\in \mathbb{R} ^{n_k\times d_{\boldsymbol{\xi }}}$ and $\dot{\boldsymbol{\varXi}}_k\in \mathbb{R} ^{n_kd_{\boldsymbol{\xi }}\times n_kd_{\boldsymbol{\xi }}}$ are the mean and covariance of the trajectory Gaussian, respectively. It can be seen that $\dot{\boldsymbol{\xi}}$ and $\dot{\boldsymbol{\varXi}}$ contain statistics of the target’s state at each time step, i.e.:
\begin{equation}
\dot{\boldsymbol{\xi}}=\left[ \left( \hat{\boldsymbol{\xi}}^1 \right) ^{\mathrm{T}},\left( \hat{\boldsymbol{\xi}}^2 \right) ^{\mathrm{T}},...,\left( \hat{\boldsymbol{\xi}}^{n_k} \right) ^{\mathrm{T}} \right] ^{\mathrm{T}}
\end{equation}
\begin{equation}
\dot{\boldsymbol{\varXi}}=\mathrm{blk}\mathrm{diag}\left( \boldsymbol{\varXi }^1,\boldsymbol{\varXi }^2,...,\boldsymbol{\varXi }^{n_k} \right)
\end{equation}

\noindent where $\hat{\boldsymbol{\xi}}^i$ and $\boldsymbol{\varXi }^i, i=1,...,n_k$ are the mean and covariance of the target's state at $i$-th time step, respectively.

To simplify the model, we make the following assumptions:
\begin{itemize}
    \item The scenario's dimension $d=2$, where the dynamic of each elliptical target follows the constant velocity model and its size does not significantly change over time.
    \item The kinematic state of the target $\boldsymbol{r},d_{o\boldsymbol{r}}=2$ includes the observable position $\boldsymbol{o}$ and the velocity $\dot{\boldsymbol{o}}$, and its shape state $\boldsymbol{s},d_{\boldsymbol{s}}=3$ includes the orientation $\boldsymbol{\theta }$ and the length of the two half axes $\boldsymbol{l}_1$ and $\boldsymbol{l}_2$. 
    \item The structured space model of the target described in Section \ref{3A} can be represented as:
    \begin{equation}
    \label{4-33}
        \boldsymbol{r}_{k}^{\left( j \right)}=\left[ \boldsymbol{o}_{\mathsf{x},k}^{\left( j \right)},\boldsymbol{o}_{\mathsf{y},k}^{\left( j \right)},\dot{\boldsymbol{o}}_{\mathsf{x},k}^{\left( j \right)},\dot{\boldsymbol{o}}_{\mathsf{y},k}^{\left( j \right)} \right] ^{\mathrm{T}}
    \end{equation}
    \begin{equation}
    \label{4-34}
        \boldsymbol{s}_{k}^{\left( j \right)}=\left[ \boldsymbol{\theta }_{k}^{\left( j \right)},\boldsymbol{l}_{1,k}^{\left( j \right)},\boldsymbol{l}_{2,k}^{\left( j \right)} \right] ^{\mathrm{T}}
    \end{equation}
    \begin{equation}
        \boldsymbol{F}_{k+1|k}^{\boldsymbol{r}}=\left[ \begin{matrix}
	1&		\mathrm{T}_{\mathrm{s}}\\
	0&		1\\
        \end{matrix} \right] \otimes \mathbf{I}_d
    \end{equation}
    \begin{equation}                \boldsymbol{F}_{k+1|k}^{\boldsymbol{s}}=\mathbf{I}_{d_{\boldsymbol{s}}}
    \end{equation}
    \begin{equation}
        \boldsymbol{F}_{k+1|k}^{\boldsymbol{\xi }}=\mathrm{blk}\mathrm{diag}\left( \boldsymbol{F}_{k+1|k}^{\boldsymbol{r}},\boldsymbol{F}_{k+1|k}^{\boldsymbol{s}} \right) 
    \end{equation}
    \begin{equation}
        \boldsymbol{Q}_{k+1|k}^{\boldsymbol{\xi }}=\mathrm{blk}\mathrm{diag}\left( \boldsymbol{Q}_{k+1|k}^{\boldsymbol{r}},\boldsymbol{Q}_{k+1|k}^{\boldsymbol{s}} \right) 
    \end{equation}
    \begin{equation}
    \boldsymbol{S}_{k}^{\left( j \right)}=\left[ \begin{matrix}
	\cos \theta _{k}^{\left( j \right)}&		-\sin \theta _{k}^{\left( j \right)}\\
	\sin \theta _{k}^{\left( j \right)}&		\cos \theta _{k}^{\left( j \right)}\\
    \end{matrix} \right] \left[ \begin{matrix}
	l_{1,k}^{\left( j \right)}&		0\\
	0&		l_{2,k}^{\left( j \right)}\\
        \end{matrix} \right] , \forall j
    \end{equation}
    \begin{equation}
        \boldsymbol{h}_{k}^{\left( j \right)}\sim \mathcal{N} \left( 0,\boldsymbol{Q}_{k}^{\boldsymbol{h}}=\frac{1}{4}\mathbf{I}_d \right) ,\forall j
    \end{equation}
    \begin{equation}
        \boldsymbol{H}_k=\left[ \left[ \begin{matrix}
	1&		0\\
\end{matrix} \right] \otimes \mathbf{I}_d,\mathbf{0}_{d\times d_{\boldsymbol{s}}} \right] 
    \end{equation}
    where the symbol `$\otimes$' represents the Kronecker product of the matrices, $\mathrm{blk}\mathrm{diag}\left( \cdot \right) $ represents constructing a block diagonal matrix. $\mathrm{T}_{\mathrm{s}}$ is the sampling interval constant.  $\mathbf{I}_a$ denotes an $a$-order identity matrix.
    \item The structured space model of the target at each time step is assumed to be invariant. For convenience, we ignore the time index, i.e., $\boldsymbol{F}_{k+1|k}^{\boldsymbol{r}}=\mathsf{F}^r$, $\boldsymbol{F}_{k+1|k}^{\boldsymbol{s}}=\mathsf{F}^{\boldsymbol{s}}$, $\boldsymbol{H}_k=\mathsf{H}$, $\boldsymbol{Q}_{k+1|k}^{\boldsymbol{r}}=\mathsf{Q}^{\boldsymbol{r}}$, $\boldsymbol{Q}_{k+1|k}^{\boldsymbol{s}}=\mathsf{Q}^{\boldsymbol{s}}$, $\boldsymbol{Q}_{k}^{\boldsymbol{h}}=\mathsf{Q}^{\boldsymbol{h}}$, $\boldsymbol{Q}_{k}^{\boldsymbol{e}}=\mathsf{Q}^{\boldsymbol{e}}$.
    \item The measurement rate of each target at each time is the same, i.e., $\gamma \left( \boldsymbol{\xi }_{k}^{\left( j \right)} \right) =\mathrm{\gamma}, \forall j,k$. And the Poisson parameter of the clutter at each time step remains constant $\lambda _k=\mathsf{\lambda}^{\mathsf{C}}, \forall k$. So given the volume of the surveillance area $\mathsf{V}$, the clutter density can be calculated as a constant $\mathsf{\rho}={{\mathsf{\lambda}^{\mathsf{C}}}/{\mathsf{V}}}$. 
    \item The detection probability $p^D\left( \cdot \right)$ and survival probability $p^S\left( \cdot \right) $ of the target are both regarded as constant $\mathsf{p}^{\mathsf{D}}$ and $\mathsf{p}^{\mathsf{S}}$.
    \item Both the PHD of the predicted density $\omega \left( \cdot \right)$, posterior density $\pi \left( \cdot \right)$, and born density $\beta \left( \cdot \right)$ follow the Gaussian mixture form, where the PHD of born density at time step $k$ can be expressed as:
    \begin{equation}
    \label{4.1.13}
        \mathscr{D} _{\beta _k}\left( \boldsymbol{\chi } \right) =\sum_{j=1}^{J_{\beta _k}}{w_{\beta _k}^{\left( j \right)}}\mathcal{N} \left( \boldsymbol{\chi };k,\hat{\boldsymbol{\xi}}_{\beta _k}^{\left( j \right)},\boldsymbol{\varXi }_{\beta _k}^{\left( j \right)} \right) 
    \end{equation}
    \begin{equation}
        n_{\beta _k}^{\left( j \right)}=1
    \end{equation}
    where $J_{\beta _k}\in \mathbb{N} $ is the number of birth components, $w_{k}^{\beta ,\left( j \right)}$ is the weight of the $j$-th birth component, and $\hat{\boldsymbol{\xi}}_{k}^{\beta ,\left( j \right)}\in \mathbb{R} ^{n_{\boldsymbol{\xi }}}$, $\boldsymbol{\varXi }_{k}^{\beta ,\left( j \right)}\in \mathbb{R} ^{n_{\boldsymbol{\xi }}\times n_{\boldsymbol{\xi }}}$ are its mean and covariance matrix, respectively. Eq.(\ref{4.1.13}) points out that each component in the Gaussian mixture is described by weight, mean, and covariance.
\end{itemize}

In the remainder of this section, we provide the Gaussian mixture implementations of the TPHD-E and TCPHD-E filters with an explicit extent update in Sections \ref{5A} and \ref{5B}, respectively.

\subsection{Gaussian Mixture TPHD Filter with Explicit Extent Update}
\label{5A}
Under the assumed model mentioned above, we can provide a closed-form recursion form of the Gaussian Mixture TPHD-E with an explicit extent update (GM-TPHD-E filter), which can be divided into the prediction step and the update step, shown in the following propositions.

\textit{Proposition 1 (GM-TPHD-E filter prediction)}: Given the PHD of the posterior multi-trajectory density $\mathscr{D} _{\pi _{k-1}}^{\mathrm{TPHD}}\left( \boldsymbol{\chi } \right) $ at time step $k-1$ as:
\begin{equation}
\mathscr{D} _{\pi _{k-1}}^{\mathsf{TPHD}}\left( \boldsymbol{\chi } \right) =\sum_{j=1}^{J_{\pi _{k-1}}}{w_{\pi _{k-1}}^{\left( j \right)}\mathcal{N} \left( \boldsymbol{\chi }; t_{\pi _{k-1}}^{\left( j \right)},\dot{\boldsymbol{\xi}}_{\pi _{k-1}}^{\left( j \right)},\dot{\boldsymbol{\varXi}}_{\pi _{k-1}}^{\left( j \right)} \right)}
\end{equation}
\noindent where $t_{\pi _{k-1}}^{\left( j \right)}+n_{\pi _{k-1}}^{\left( j \right)}-1=k-1$ with $n_{\pi _{k-1}}^{\left( j \right)}={{\mathrm{dim}( \dot{\boldsymbol{\xi}}_{\pi _{k-1}}^{\left( j \right)})}/{d_{\boldsymbol{\xi }}}}$. And $J_{\pi _{k-1}}\in \mathbb{N} $ is the number of components contained in $\mathscr{D} _{\pi _{k-1}}^{\mathsf{TPHD}}\left( \cdot \right)$.

Then the PHD of predicted multi-trajectory density $\mathscr{D} _{\omega _k}^{\mathsf{TPHD}}\left( \boldsymbol{\chi } \right)$ at time step $k$ is: 
\begin{equation}
\mathscr{D} _{\omega _k}^{\mathsf{TPHD}}\left( \boldsymbol{\chi } \right) =\mathscr{D} _{\beta _k}+\mathsf{p}^{\mathsf{S}}\sum_{j=1}^{J_{\pi _{k-1}}}{w_{\pi _{k-1}}^{\left( j \right)}\mathcal{N} \left( \boldsymbol{\chi };t_{\pi _{k-1}}^{\left( j \right)},\dot{\boldsymbol{\xi}}_{\omega _k}^{\left( j \right)},\dot{\boldsymbol{\varXi}}_{\omega _k}^{\left( j \right)} \right)}
\end{equation}
\noindent where
\begin{equation}
w_{\omega _k}^{\left( j \right)}=w_{\pi _{k-1}}^{\left( j \right)}
\end{equation}
\begin{equation}
\label{4.1.18}
\dot{\boldsymbol{\xi}}_{\omega _{k-1}}^{\left( j \right)}=\left[ \left( \dot{\boldsymbol{\xi}}_{\pi _{k-1}}^{\left( j \right)} \right) ^{\mathrm{T}},\left( \dot{\boldsymbol{F}}_{\pi _{k-1}}^{\boldsymbol{\xi },\left( j \right)}\dot{\boldsymbol{\xi}}_{\pi _{k-1}}^{\left( j \right)} \right) ^{\mathrm{T}} \right] ^{\mathrm{T}}
\end{equation}
\begin{equation}
\label{4.1.19}
\dot{\boldsymbol{\varXi}}_{\omega _{k-1}}^{\left( j \right)}=\left[ \begin{matrix}
	\dot{\boldsymbol{\varXi}}_{\pi _{k-1}}^{\left( j \right)}&		\dot{\boldsymbol{\varXi}}_{\pi _{k-1}}^{\left( j \right)}\left( \dot{\boldsymbol{F}}_{\pi _{k-1}}^{\boldsymbol{\xi },\left( j \right)} \right) ^{\mathrm{T}}\\
	\dot{\boldsymbol{F}}_{\pi _{k-1}}^{\boldsymbol{\xi },\left( j \right)}\dot{\boldsymbol{\varXi}}_{\pi _{k-1}}^{\left( j \right)}&		\dot{\boldsymbol{F}}_{\pi _{k-1}}^{\boldsymbol{\xi },\left( j \right)}\dot{\boldsymbol{\varXi}}_{\pi _{k-1}}^{\left( j \right)}\left( \dot{\boldsymbol{F}}_{\pi _{k-1}}^{\boldsymbol{\xi },\left( j \right)} \right) ^{\mathrm{T}}+\mathsf{Q}^{\boldsymbol{\xi }}\\
\end{matrix} \right] 
\end{equation}
\begin{equation}
\label{4.1.20}
\dot{\boldsymbol{F}}_{\pi _{k-1}}^{\boldsymbol{\xi },\left( j \right)}=\left[ \mathbf{0}_{1,n_{\pi _{k-1}}^{\left( j \right)}},1 \right] \otimes \mathsf{F}^{\boldsymbol{\xi }}
\end{equation}
\begin{equation}
t_{\omega _k}^{\left( j \right)}=t_{\pi _{k-1}}^{\left( j \right)}
\end{equation}

Note that we use the superscript `$\mathsf{TPHD}$' to indicate variables related to the TPHD filter, and `$\mathbf{0}_{a,b}$' in Eq.(\ref{4.1.20}) represents a zero matrix of size $a\times b$. It can be seen that in the prediction step, the length of each alive trajectory is extended by one time step, i.e.:
\begin{equation}
n_{\omega _k}^{\left( j \right)}=n_{\pi _{k-1}}^{\left( j \right)}+1
\end{equation}

\textit{Proposition 2 (GM-TPHD-E filter update)}: Given the PHD of the predicted multi-trajectory density $\mathscr{D} _{\omega _k}^{\mathsf{TPHD}}\left( \boldsymbol{\chi } \right) $ at time step $k$ as:
\begin{equation}
\mathscr{D} _{\omega _k}^{\mathsf{TPHD}}\left( \boldsymbol{\chi } \right) =\sum_{j=1}^{J_{\omega _k}}{w_{\omega _k}^{\left( j \right)}\mathcal{N} \left( \boldsymbol{\chi }; t_{\omega _k}^{\left( j \right)},\dot{\boldsymbol{\xi}}_{\omega _k}^{\left( j \right)},\dot{\boldsymbol{\varXi}}_{\omega _k}^{\left( j \right)} \right)}
\end{equation}
\noindent where $t_{\omega _k}^{\left( j \right)}+n_{\omega _k}^{\left( j \right)}-1=k$ with $n_{\omega _k}^{\left( j \right)}={{\mathrm{dim}( \dot{\boldsymbol{\xi}}_{\omega _k}^{\left( j \right)})}/{d_{\boldsymbol{\xi }}}}$. And $J_{\omega _k}\in \mathbb{N} $ is the number of components in $\mathscr{D} _{\omega _k}^{\mathsf{TPHD}}\left( \cdot \right)$, satisfying $J_{\omega _k}=J_{\pi _{k-1}}+J_{\beta _k}$.

Then the PHD of the posterior multi-trajectory density $\mathscr{D} _{\pi _k}^{\mathsf{TPHD}}\left( \boldsymbol{\chi } \right) $ with the extent update at time step $k$ is:
\begin{equation}
\begin{split}
&\mathscr{D} _{\pi _k}^{\mathsf{TPHD}}\left( \boldsymbol{\chi } \right) \\
&=\left( 1-\left( 1-\mathrm{e}^{-\gamma} \right) \mathsf{p}^{\mathsf{D}} \right) \mathscr{D} _{\omega _k}^{\mathsf{TPHD}}\left( \boldsymbol{\chi } \right) +\sum_{\mathcal{P} \angle \mathcal{Z} _k}{\sum_{\mathcal{C} \in \mathcal{P}}{\mathscr{D} _{\pi _k}^{\mathsf{D}}\left( \boldsymbol{\chi },\mathcal{P} ,\mathcal{C} \right)}}
\end{split}
\end{equation}

\begin{equation}
\mathscr{D} _{\pi _k}^{\mathsf{D}}\left( \boldsymbol{\chi },\mathcal{P} ,\mathcal{C} \right) =\sum_{j=1}^{J_{\omega _k}}{w_{\pi _k}^{\left( j,\mathcal{P} ,\mathcal{C} \right)}\mathcal{N} \left( \boldsymbol{\chi };t_{\pi _k}^{\left( j,\mathcal{P} ,\mathcal{C} \right)},\dot{\boldsymbol{\xi}}_{\pi _k}^{\left( j,\mathcal{P} ,\mathcal{C} \right)},\dot{\boldsymbol{\varXi}}_{\pi _k}^{\left( j,\mathcal{P} ,\mathcal{C} \right)} \right)}
\end{equation}
where
\begin{equation}
\label{4.1.25}
w_{\pi _k}^{\left( j,\mathcal{P} ,\mathcal{C} \right)}=\mathsf{p}^{\mathsf{D}}\frac{\mathsf{e}^{-\mathrm{\gamma}}w_{\mathcal{P}}}{\left( \varrho _{\left( \mathcal{P} ,\mathcal{C} \right)}+\varsigma _{\left( \mathcal{P} ,\mathcal{C} \right)} \right)}\mathrm{\gamma}^{\left| \mathcal{P} ,\mathcal{C} \right|}\mathcal{L} _{w_{k}^{\tau}}^{\left( j,\mathcal{P} ,\mathcal{C} \right)}w_{\omega _k}^{\left( j \right)}
\end{equation}
\begin{equation}
\label{4.1.26}
\dot{\boldsymbol{\xi}}_{\pi _k}^{\left( j,\mathcal{P} ,\mathcal{C} \right)}=\left[ \mathscr{I} \left( n_{\omega _k}^{\left( j \right)} \right) \dot{\boldsymbol{\xi}}_{\omega _k}^{\left( j \right)}, \left( \hat{\boldsymbol{\xi}}_{\pi _k}^{\left( j,\mathcal{P} ,\mathcal{C} \right)} \right) ^{\mathrm{T}} \right] ^{\mathrm{T}}
\end{equation}
\begin{equation}
\label{4.1.27}
\dot{\boldsymbol{\varXi}}_{\pi _k}^{\left( j,\mathcal{P} ,\mathcal{C} \right)}=\mathrm{blk}\mathrm{diag}\left( \mathscr{I} \left( n_{\omega _k}^{\left( j \right)} \right) \dot{\boldsymbol{\varXi}}_{\omega _k}^{\left( j \right)}\left( \mathscr{I} \left( n_{\omega _k}^{\left( j \right)} \right) \right) ^{\mathrm{T}},\boldsymbol{\varXi }_{\pi _k}^{\left( j,\mathcal{P} ,\mathcal{C} \right)} \right) 
\end{equation}
\begin{equation}
\label{4.1.28}
\mathscr{I} \left( n_{\omega _k}^{\left( j \right)} \right) =\left[ \mathbf{I}_{\left( n_{\omega _k}^{\left( j \right)}-1 \right) d_{\boldsymbol{\xi }}},\mathbf{O}_{d_{\boldsymbol{\xi }}}\otimes \left( \mathbf{1}_{1,n_{\omega _k}^{\left( j \right)}-1} \right) ^{\mathrm{T}} \right]  
\end{equation}
\begin{equation}
\label{4.1.29}
t_{\pi _k}^{\left( j,\mathcal{P} ,\mathcal{C} \right)}=t_{\omega _k}^{\left( j \right)}
\end{equation}

In Eq.(\ref{4.1.25}), $\mathcal{L} _{\pi _{k}^{\tau}}^{\left( j,\mathcal{P} ,\mathcal{C} \right)}$ represents the measurement likelihood of the measurement cell $\left( \mathcal{P} ,\mathcal{C} \right)$ condition on the target state corresponding to the $j$-th component of $\mathscr{D} _{\omega _k}^{\mathsf{TPHD}}\left( \boldsymbol{\chi } \right) $. Given the predicted trajectory component $\mathcal{N} \left( \boldsymbol{\chi }; t_{\omega _k}^{\left( j \right)},\dot{\boldsymbol{\xi}}_{\omega _k}^{\left( j \right)},\dot{\boldsymbol{\varXi}}_{\omega _k}^{\left( j \right)} \right) $ with length $n_{\omega _k}^{\left( j \right)}$, its corresponding target state $\hat{\boldsymbol{\xi}}_{\omega _k}^{\left( j \right)}$ and covariance matrix $\boldsymbol{\varXi }_{\omega _k}^{\left( j \right)}$ are:
\begin{equation}
\begin{split}
\hat{\boldsymbol{\xi}}_{\omega _k}^{\left( j \right)}&=\mathscr{O} \left( n_{\omega _k}^{\left( j \right)} \right) \dot{\boldsymbol{\xi}}_{\omega _k}^{\left( j \right)}=\left[ \left( \hat{\boldsymbol{r}}_{\omega _k}^{\left( j \right)} \right) ^{\mathrm{T}},\left( \hat{\boldsymbol{s}}_{\omega _k}^{\left( j \right)} \right) ^{\mathrm{T}} \right] ^{\mathrm{T}}
\\
&=\left[ \left( \hat{\boldsymbol{r}}_{\omega _k}^{\left( j \right)} \right) ^{\mathrm{T}},\left( \hat{\theta}_{\omega _k}^{\left( j \right)}, \hat{l}_{1,\omega _k}^{\left( j \right)}, \hat{l}_{2,\omega _k}^{\left( j \right)} \right) ^{\mathrm{T}} \right] ^{\mathrm{T}}
\end{split}
\end{equation}
\begin{equation}
\begin{split}
\boldsymbol{\varXi }_{\omega _k}^{\left( j \right)}&=\mathscr{O} \left( n_{\omega _k}^{\left( j \right)} \right) \dot{\boldsymbol{\varXi}}_{\omega _k}^{\left( j \right)}\left( \mathscr{O} \left( n_{\omega _k}^{\left( j \right)} \right) \right) ^{\mathrm{T}}
\\
&=\mathrm{blk}\mathrm{diag}\left( \boldsymbol{\varXi }_{\omega _k}^{\boldsymbol{r}\left( j \right)}, \boldsymbol{\varXi }_{\omega _k}^{\boldsymbol{s}\left( j \right)} \right) 
\end{split}
\end{equation}
\begin{equation}
\mathscr{O} \left( n_{\omega _k}^{\left( j \right)} \right) =\left[ \mathbf{O}_{d_{\boldsymbol{\xi }}}\otimes \mathbf{1}_{1,n_{\omega _k}^{\left( j \right)}-1},\mathbf{I}_{d_{\boldsymbol{\xi }}} \right] 
\end{equation}

Then the measurement likelihood $\mathcal{L} _{w_{k}^{\tau}}^{\left( j,\mathcal{P} ,\mathcal{C} \right)}$ can be calculated as:
\begin{equation}
\label{4.1.33}
\begin{split}
\mathcal{L} _{w_{k}^{\tau}}^{\left( j,\mathcal{P} ,\mathcal{C} \right)}&=\prod_{\boldsymbol{z}_k\in \left( \mathcal{P} ,\mathcal{C} \right)}{\varphi _{\boldsymbol{z}_k}\left( \hat{\boldsymbol{\xi}}_{\omega _k}^{\left( j \right)} \right)} \\
&=\prod_{m=1}^{\left| \mathcal{P} ,\mathcal{C} \right|}{\mathcal{N} \left( \boldsymbol{z}_{k}^{\mathcal{P} ,\mathcal{C} ,m};\mathsf{H}\hat{\boldsymbol{r}}_{\omega _k}^{\left( j \right)},\boldsymbol{\varXi }_{\omega _k}^{\boldsymbol{z}\left( j \right)} \right)}
\end{split}
\end{equation}

\begin{equation}
\boldsymbol{\varXi }_{\omega _k}^{\boldsymbol{z}\left( j \right)}=\mathsf{H}\boldsymbol{\varXi }_{\omega _k}^{\boldsymbol{r}\left( j \right)}\mathsf{H}^{\mathrm{T}}+\boldsymbol{S}_{\omega _k}^{\left( j \right)}\mathsf{Q}^{\boldsymbol{h}}\left( \boldsymbol{S}_{\omega _k}^{\left( j \right)} \right) ^{\mathrm{T}}+\mathsf{Q}^{\boldsymbol{e}}
\end{equation}

\begin{equation}
\boldsymbol{S}_{\omega _k}^{\left( j \right)}=\left[ \begin{matrix}
	\cos \hat{\theta}_{\omega _k}^{\left( j \right)}&		-\sin \hat{\theta}_{\omega _k}^{\left( j \right)}\\
	\sin \hat{\theta}_{\omega _k}^{\left( j \right)}&		\cos \hat{\theta}_{\omega _k}^{\left( j \right)}\\
\end{matrix} \right] \left[ \begin{matrix}
	\hat{l}_{1,\omega _k}^{\left( j \right)}&		0\\
	0&		\hat{l}_{2,\omega _k}^{\left( j \right)}\\
\end{matrix} \right] 
\end{equation}

\noindent where $\boldsymbol{z}_{k}^{\mathcal{P} ,\mathcal{C} ,m}$ in Eq.(\ref{4.1.33}) represents the $m$-th measurement in the cell $\left( \mathcal{P} ,\mathcal{C} \right) $, and we denote all measurements contained in cell $\left( \mathcal{P} ,\mathcal{C} \right) $ as $\left\{ \boldsymbol{z}_{k}^{\mathcal{P} ,\mathcal{C} ,m} \right\} _{m=1}^{\left| \mathcal{P} ,\mathcal{C} \right|}$. 

The coefficients $\varrho _{\left( \mathcal{P} ,\mathcal{C} \right)}$ and $\varsigma _{\left( \mathcal{P} ,\mathcal{C} \right)}$ in Eq.(\ref{4.1.25}) can be obtained based on Eq.(\ref{3.1.6}) and Eq.(\ref{3.1.7}), respectively. Here we have:
\begin{equation}
\varsigma _{\left( \mathcal{P} ,\mathcal{C} \right)}=\sum_{n=1}^{J_{\omega _k}}{\mathsf{e}^{-\mathrm{\gamma}}\mathrm{\gamma}^{\left| \mathrm{p},\mathcal{C} \right|}\mathsf{p}^{\mathsf{D}}}\mathcal{L} _{\pi _{k}^{\tau}}^{\left( j,\mathcal{P} ,\mathcal{C} \right)}w_{\omega _k}^{\left( j \right)}
\end{equation}
\begin{equation}
\varrho _{\left( \mathcal{P} ,\mathcal{C} \right)}=\delta _{\left| \mathcal{P} ,\mathcal{C} \right|,1}\mathrm{\rho}^{\left| \mathcal{P} ,\mathcal{C} \right|}
\end{equation}

Then the weight $\omega _{\mathcal{P}}$ of the partition $\mathcal{P} $ in Eq.(\ref{4.1.25}) can be calculated as Eq.(\ref{3.1.8}).

In addition, the update target state $\hat{\boldsymbol{\xi}}_{\pi _k}^{\left( j,\mathcal{P} ,\mathcal{C} \right)}$ and the corresponding covariance matrix $\boldsymbol{\varXi }_{\pi _k}^{\left( j,\mathcal{P} ,\mathcal{C} \right)}$ in Eq.(\ref{4.1.26}) and Eq.(\ref{4.1.27}) are obtained by sequentially measurement updating the predicted target state $\hat{\boldsymbol{\xi}}_{\omega _k}^{\left( j \right)}$ and the covariance matrix using the measurement cell $\left( \mathcal{P} ,\mathcal{C} \right)$. Specifically, the sequential measurement update refers to recursively updating the state of the target in a non-repeating order of measurements. That is, given the update state $\hat{\boldsymbol{\xi}}_{\pi _k,m}^{\left( j,\mathcal{P} ,\mathcal{C} \right)}=\left[ \left( \hat{\boldsymbol{r}}_{\pi _k,m}^{\left( j,\mathrm{p},\mathcal{C} \right)} \right) ^{\mathrm{T}},\left( \hat{\boldsymbol{s}}_{\pi _k,m}^{\left( j,\mathrm{p},\mathcal{C} \right)} \right) ^{\mathrm{T}} \right] ^{\mathrm{T}}$ and the corresponding covariance $\boldsymbol{\varXi }_{\pi _k,m}^{\left( j,\mathcal{P} ,\mathcal{C} \right)}=\mathrm{blk}\mathrm{diag}\left( \boldsymbol{\varXi }_{\pi _k,m}^{\boldsymbol{r}\left( j,\mathrm{p},\mathcal{C} \right)},\boldsymbol{\varXi }_{\pi _k,m}^{\boldsymbol{s}\left( j,\mathrm{p},\mathcal{C} \right)} \right) $, they indicate the estimate and covariance of the $j$-th target incorporated up to the $m$-th measurement $\boldsymbol{z}_{k}^{\mathcal{P} ,\mathcal{C} ,m}$ in the cell $\left( \mathcal{P} ,\mathcal{C} \right) $. 

Now, given the $m+1$-th measurement $\boldsymbol{z}_{k}^{\mathcal{P} ,\mathcal{C} ,m+1}$, ignore the subscript `$\pi _k$' representing the posterior of time step $k$ and the superscripts of the measurement cell `$\left( \mathcal{P} ,\mathcal{C} \right) $', i.e., we abbreviate the measurement $\boldsymbol{z}_{k}^{\mathcal{P} ,\mathcal{C} ,m+1}$ as $\boldsymbol{z}^{m+1}$, then the updated target state $\hat{\boldsymbol{\xi}}_{\pi _k,m}^{\left( j,\mathcal{P} ,\mathcal{C} \right)}$ and the corresponding covariance matrix $\boldsymbol{\varXi }_{\pi _k,m}^{\left( j,\mathcal{P} ,\mathcal{C} \right)}$ are simply represented as $\hat{\boldsymbol{\xi}}_{m}^{\left( j \right)}$ and $\boldsymbol{\varXi }_{m}^{\left( j \right)}$, respectively, calculated as:
\begin{equation}
\hat{\boldsymbol{\xi}}_{m+1}^{\left( j \right)}=\hat{\boldsymbol{\xi}}_{m}^{\left( j \right)}+\boldsymbol{\varXi }_{m+1}^{\boldsymbol{\xi }\bar{\boldsymbol{Z}}\left( j \right)}\left( \boldsymbol{\varXi }_{m+1}^{\bar{\boldsymbol{Z}}\left( j \right)} \right) ^{-1}\left( \bar{\boldsymbol{Z}}_{m+1}-\tilde{\boldsymbol{Z}}_{m+1} \right) 
\end{equation}
\begin{equation}
\boldsymbol{\varXi }_{m+1}^{\left( j \right)}=\boldsymbol{\varXi }_{m}^{\left( j \right)}-\boldsymbol{\varXi }_{m+1}^{\boldsymbol{\xi }\bar{\boldsymbol{Z}}\left( j \right)}\left( \boldsymbol{\varXi }_{m+1}^{\bar{\boldsymbol{Z}}\left( j \right)} \right) ^{-1}\left( \boldsymbol{\varXi }_{m+1}^{\boldsymbol{\xi }\bar{\boldsymbol{Z}}\left( j \right)} \right) ^{\mathrm{T}}
\end{equation}
\begin{equation}
\boldsymbol{\varXi }_{m+1}^{\bar{\boldsymbol{Z}}\left( j \right)}=\left[ \begin{matrix}
	\boldsymbol{\varXi }_{m+1}^{\boldsymbol{z}\left( j \right)}&		0\\
	0&		\boldsymbol{\varXi }_{m+1}^{\boldsymbol{Z}\left( j \right)}\\
\end{matrix} \right] \,\,\mathrm{and} \,\, \boldsymbol{\varXi }_{m+1}^{\boldsymbol{\xi }\bar{\boldsymbol{Z}}\left( j \right)}=\left[ \begin{matrix}
	\boldsymbol{\varXi }_{m+1}^{\boldsymbol{rz}\left( j \right)}&		0\\
	0&		\boldsymbol{\varXi }_{m+1}^{\boldsymbol{sz}\left( j \right)}\\
\end{matrix} \right] 
\end{equation}
\begin{equation}
\bar{\boldsymbol{Z}}_{m+1}=\left[ \begin{array}{c}
	\boldsymbol{z}^{m+1}\\
	\boldsymbol{Z}^{m+1}\\
\end{array} \right] \,\, \mathrm{and} \,\, \tilde{\boldsymbol{Z}}_{m+1}=\left[ \begin{array}{c}
	\bar{\boldsymbol{z}}^{m+1}\\
	\bar{\boldsymbol{Z}}^{m+1}\\
\end{array} \right] 
\end{equation}
\noindent where
\begin{equation}
\label{4.1.39}
\begin{split}
\boldsymbol{\varXi }_{m+1}^{\boldsymbol{z}\left( j \right)}&=\mathsf{H}\boldsymbol{\varXi }_{m}^{\boldsymbol{r}\left( j \right)}\mathsf{H}^{\mathrm{T}}+\boldsymbol{S}_{m}^{\left( j \right)}\mathsf{Q}^{\boldsymbol{h}}\left( \boldsymbol{S}_{m}^{\left( j \right)} \right) ^{\mathrm{T}}+\mathsf{Q}^{\boldsymbol{e}}\\
&=\left[ \begin{matrix}
	\sigma _{11}&		\sigma _{12}\\
	\sigma _{21}&		\sigma _{22}\\
\end{matrix} \right] 
\end{split}
\end{equation}
\begin{equation}
\boldsymbol{S}_{m}^{\left( j \right)}=\left[ \begin{matrix}
	\cos \hat{\theta}_{m}^{\left( j \right)}&		-\sin \hat{\theta}_{m}^{\left( j \right)}\\
	\sin \hat{\theta}_{m}^{\left( j \right)}&		\cos \hat{\theta}_{m}^{\left( j \right)}\\
\end{matrix} \right] \left[ \begin{matrix}
	\hat{l}_{1,m}^{\left( j \right)}&		0\\
	0&		\hat{l}_{2,m}^{\left( j \right)}\\
\end{matrix} \right] 
\end{equation}
\begin{equation}
\boldsymbol{\varXi }_{m+1}^{\boldsymbol{rz}\left( j \right)}=\boldsymbol{\varXi }_{m}^{\boldsymbol{r}\left( j \right)}\mathsf{H}^{\mathrm{T}}
\end{equation}
\begin{equation}
\boldsymbol{\varXi }_{m+1}^{\boldsymbol{sz}\left( j \right)}=\boldsymbol{\varXi }_{m}^{\boldsymbol{s}\left( j \right)}\left( \boldsymbol{Y}_{m}^{\left( j \right)} \right) ^{\mathrm{T}}
\end{equation}
\begin{equation}
\boldsymbol{Y}_{m}^{\left( j \right)}=\left[ \begin{array}{c}
	2\boldsymbol{S}_{1,m}^{\left( j \right)}\mathsf{Q}^h\boldsymbol{J}_{1,m}^{\left( j \right)}\\
	2\boldsymbol{S}_{2,m}^{\left( j \right)}\mathsf{Q}^h\boldsymbol{J}_{2,m}^{\left( j \right)}\\
	\boldsymbol{S}_{1,m}^{\left( j \right)}\mathsf{Q}^h\boldsymbol{J}_{2,m}^{\left( j \right)}+\boldsymbol{S}_{2,m}^{\left( j \right)}\mathsf{Q}^h\boldsymbol{J}_{1,m}^{\left( j \right)}\\
\end{array} \right] 
\end{equation}
\begin{equation}
\boldsymbol{S}_{1,m}^{\left( j \right)}=\left[ \hat{l}_{1,m}^{\left( j \right)}\cos \hat{\theta}_{m}^{\left( j \right)},-\hat{l}_{2,m}^{\left( j \right)}\sin \hat{\theta}_{m}^{\left( j \right)} \right] 
\end{equation}
\begin{equation}
\boldsymbol{S}_{2,m}^{\left( j \right)}=\left[ \hat{l}_{1,m}^{\left( j \right)}\sin \hat{\theta}_{m}^{\left( j \right)},\hat{l}_{2,m}^{\left( j \right)}\cos \hat{\theta}_{m}^{\left( j \right)} \right] 
\end{equation}
\begin{equation}
\boldsymbol{J}_{1,m}^{\left( j \right)}=\left[ \begin{array}{c}
	-\hat{l}_{1,m}^{\left( j \right)}\sin \hat{\theta}_{m}^{\left( j \right)}\\
	-\hat{l}_{2,m}^{\left( j \right)}\cos \hat{\theta}_{m}^{\left( j \right)}\\
\end{array}\begin{array}{c}
	\cos \hat{\theta}_{m}^{\left( j \right)}\\
	0\\
\end{array}\begin{array}{c}
	0\\
	-\sin \hat{\theta}_{m}^{\left( j \right)}\\
\end{array} \right] 
\end{equation}
\begin{equation}
\boldsymbol{J}_{2,m}^{\left( j \right)}=\left[ \begin{array}{c}
	\hat{l}_{1,m}^{\left( j \right)}\cos \hat{\theta}_{m}^{\left( j \right)}\\
	-\hat{l}_{2,m}^{\left( j \right)}\sin \hat{\theta}_{m}^{\left( j \right)}\\
\end{array}\begin{array}{c}
	\sin \hat{\theta}_{m}^{\left( j \right)}\\
	0\\
\end{array}\begin{array}{c}
	0\\
	\cos \hat{\theta}_{m}^{\left( j \right)}\\
\end{array} \right] 
\end{equation}
\begin{equation}
\boldsymbol{Z}^{m+1}=\mathring{\mathsf{F}}\left( \left( \boldsymbol{z}^{m+1}-\bar{\boldsymbol{z}}^{m+1} \right) \otimes \left( \boldsymbol{z}^{m+1}-\bar{\boldsymbol{z}}^{m+1} \right) \right) 
\end{equation}
\begin{equation}
\mathring{\mathsf{F}}=\left[ \begin{array}{c}
	\begin{matrix}
	1&		0&		0&		0\\
\end{matrix}\\
	\begin{matrix}
	0&		0&		0&		1\\
\end{matrix}\\
	\begin{matrix}
	0&		1&		0&		0\\
\end{matrix}\\
\end{array} \right]     
\end{equation}
\begin{equation}
\bar{\boldsymbol{z}}^{m+1}=\mathrm{H}\hat{\boldsymbol{r}}_{m}^{\left( j \right)}
\end{equation}
\begin{equation}
\bar{\boldsymbol{Z}}^{m+1}=\left[ \begin{matrix}
	\sigma _{11}&		\sigma _{22}&		\sigma _{12}\\
\end{matrix} \right] ^{\mathrm{T}}
\end{equation}
\begin{equation}
\label{4.1.51}
\boldsymbol{\varXi }_{m+1}^{\boldsymbol{Z}\left( j \right)}=\left[ \begin{matrix}
	2{\sigma _{11}}^2&		2{\sigma _{12}}^2&		2\sigma _{11}\sigma _{12}\\
	2{\sigma _{12}}^2&		2{\sigma _{22}}^2&		2\sigma _{22}\sigma _{12}\\
	2\sigma _{11}\sigma _{12}&		2\sigma _{12}\sigma _{22}&		\sigma _{11}\sigma _{22}+{\sigma _{12}}^2\\
\end{matrix} \right] 
\end{equation}

Refer to \cite{2019YangTSP} for details on the derivation of Eq.(\ref{4.1.39})$\sim$Eq.(\ref{4.1.51}). Note that in Eq.(\ref{4.1.26}) and Eq.(\ref{4.1.27}), we construct $\dot{\boldsymbol{\xi}}_{\pi _k}^{\left( j,\mathcal{P} ,\mathcal{C} \right)}$ and $\dot{\boldsymbol{\varXi}}_{\pi _k}^{\left( j,\mathcal{P} ,\mathcal{C} \right)}$ through matrix concatenation, which has a slight error because it assumes that the target states at different time steps do not affect each other. In fact, such an error can be ignored, and this construction method is similar to the ``1-Scan'' implementation in \cite{2019GarcíaTSP}.

\subsection{Gaussian Mixture TCPHD-E Filter with Explicit Extent Update}
\label{5B}
Similar to the GM-TPHD-E filter, the Gaussian Mixture TCPHD-E filter with an explicit extent update (GM-TCPHD-E filter) also has a recursive process of prediction and update, and simultaneously propagates multi-trajectory density and cardinality distribution, represented in the following propositions.

\textit{Proposition 3 (GM-TCPHD-E filter prediction)}:  Given the cardinality distribution $\mathscr{P} _{\pi _{k-1}}\left( n \right)$ and the PHD of posterior multi-trajectory density $\mathscr{D} _{\pi _{k-1}}^{\mathsf{TCPHD}}\left( \boldsymbol{\chi } \right) $ at time step $k-1$ as:
\begin{equation}
\mathscr{D} _{\pi _{k-1}}^{\mathsf{TCPHD}}\left( \boldsymbol{\chi }_{k-1} \right) =\sum_{j=1}^{J_{\pi _{k-1}}}{w_{\pi _{k-1}}^{\left( j \right)}\mathcal{N} \left( \boldsymbol{\chi };t_{\pi _{k-1}}^{\left( j \right)},\dot{\boldsymbol{\xi}}_{\pi _{k-1}}^{\left( j \right)},\dot{\boldsymbol{\varXi}}_{\pi _{k-1}}^{\left( j \right)} \right)}
\end{equation}

Then, the GM-TCPHD-E filter prediction yields the predicted cardinality distribution $\mathscr{P} _{\omega _k}\left( \cdot \right) $ and the PHD of predicted multi-trajectory density $\mathscr{D} _{\omega _k}^{\mathsf{TCPHD}}\left( \cdot \right) $, denoted as:
\begin{equation}
\begin{split}
&\mathscr{P} _{\omega _k}\left( n \right)= \\
&\sum_{j=0}^n{\left[ \mathscr{P} _{\beta _{k}^{\tau}}\left( n-j \right) \sum_{l-j}^{\infty}{\left( \binom{l}{j}  \mathscr{P} _{\pi _{k-1}}\left( n \right) \left( \mathsf{p}^{\mathsf{S}} \right) ^j\left( 1-\mathsf{p}^{\mathsf{S}} \right) ^{l-j} \right)} \right]}
\end{split}
\end{equation}

\begin{equation}
\mathscr{D} _{\omega _k}^{\mathsf{TCPHD}}\left( \boldsymbol{\chi } \right) =\mathscr{D} _{\beta _k}+\mathsf{p}^{\mathsf{S}}\sum_{j=1}^{J_{\pi _{k-1}}}{w_{\pi _{k-1}}^{\left( j \right)}\mathcal{N} \left( \boldsymbol{\chi };t_{\pi _{k-1}}^{\left( j \right)},\dot{\boldsymbol{\xi}}_{\omega _k}^{\left( j \right)},\dot{\boldsymbol{\varXi}}_{\omega _k}^{\left( j \right)} \right)}
\end{equation}
\noindent where $\dot{\boldsymbol{\xi}}_{\omega _k}^{\left( j \right)}$ and $\dot{\boldsymbol{\varXi}}_{\omega _k}^{\left( j \right)}$ are given in Eq.(\ref{4.1.18}) and Eq.(\ref{4.1.19}), respectively.

\textit{Proposition 4 (GM-TCPHD-E filter update)}: Given the predicted cardinality distribution $\mathscr{P} _{\omega _k}\left( n \right)$ and the PHD of the predicted multi-trajectory density $\mathscr{D} _{\omega _k}^{\mathsf{TCPHD}}\left( \boldsymbol{\chi } \right)$ at time step $k$ as:
\begin{equation}
\mathscr{D} _{\omega _k}^{\mathsf{TCPHD}}\left( \boldsymbol{\chi } \right) =\sum_{j=1}^{J_{\omega _k}}{w_{\omega _k}^{\left( j \right)}\mathcal{N} \left( \boldsymbol{\chi };t_{\omega _k}^{\left( j \right)},\dot{\boldsymbol{\xi}}_{\omega _k}^{\left( j \right)},\dot{\boldsymbol{\varXi}}_{\omega _k}^{\left( j \right)} \right)}
\end{equation}

Then, the GM-TCPHD-E filter update yields the cardinality distribution  $\mathscr{P} _{\pi _k}\left( \cdot \right)$ and the PHD of posterior multi-trajectory density $\mathscr{D} _{\pi _k}^{\mathrm{TCPHD}}\left( \cdot \right) $, denoted as:

\begin{equation}
\begin{split}
\mathscr{D} _{\pi _k}^{\mathsf{TCPHD}}\left( \boldsymbol{\chi } \right) =&\kappa \left( 1-\left( 1-\mathsf{e}^{-\gamma} \right) \mathsf{p}^{\mathsf{D}} \right) \mathscr{D} _{\omega _k}^{\mathsf{TCPHD}}\left( \boldsymbol{\chi } \right) \\
&+\sum_{\mathcal{P} \angle \mathcal{Z} _k}{\sum_{\mathcal{C} \in \mathcal{P}}{\mathscr{D} _{\pi _k}^{\mathsf{D}^{\prime}}\left( \boldsymbol{\chi },\mathcal{P} ,\mathcal{C} \right)}}
\end{split}
\end{equation}
\begin{equation}
\mathscr{D} _{\pi _k}^{\mathsf{D}^{\prime}}\left( \boldsymbol{\chi },\mathcal{P} ,\mathcal{C} \right) =\sum_{j=1}^{J_{\omega _k}}{w_{\pi _k}^{\left( j,\mathcal{P} ,\mathcal{C} \right)}\mathcal{N} \left( \boldsymbol{\chi };t_{\pi _k}^{\left( j,\mathcal{P} ,\mathcal{C} \right)},\dot{\boldsymbol{\xi}}_{\pi _k}^{\left( j,\mathcal{P} ,\mathcal{C} \right)},\dot{\boldsymbol{\varXi}}_{\pi _k}^{\left( j,\mathcal{P} ,\mathcal{C} \right)} \right)}
\end{equation}
\noindent where
\begin{equation}
\label{4.2.7}
w_{\pi _k}^{\left( j,\mathcal{P} ,\mathcal{C} \right)}=\frac{\mathsf{p}^{\mathsf{D}}\bar{w}_{\omega _k}^{\left( j \right)}\nu _{\mathcal{P} ,\mathcal{C}}\frac{\mathcal{L} _{w_{k}^{\tau}}^{\left( j,\mathcal{P} ,\mathcal{C} \right)}}{\mathrm{\rho}^{\left| \mathcal{P} ,\mathcal{C} \right|}}}{\sum_{\mathcal{P} \angle \mathcal{Z} _k}{\sum_{\mathcal{C} \in \mathcal{P}}{\vartheta _{\mathcal{P} ,\mathcal{C}}\epsilon _{\mathcal{P} ,\mathcal{C}}}}}
\end{equation}
\begin{equation}
\bar{w}_{\omega _k}^{\left( j \right)}\triangleq \frac{w_{\omega _k}^{\left( j \right)}}{\sum\nolimits_{l=1}^{J_{\omega _k}}{w_{\omega _k}^{\left( l \right)}}}
\end{equation}

The calculation of $\dot{\boldsymbol{\xi}}_{\pi _k}^{\left( j,\mathcal{P} ,\mathcal{C} \right)}$ and $\dot{\boldsymbol{\varXi}}_{\pi _k}^{\left( j,\mathcal{P} ,\mathcal{C} \right)}$ are given in Proposition 2. And the measurement likelihood  $\mathcal{L} _{w_{k}^{\tau}}^{\left( j,\mathcal{P} ,\mathcal{C} \right)}$ in Eq.(\ref{4.2.7}) can be gained according to Eq.(\ref{4.1.33}). Moreover, the coefficients $\kappa$, $\vartheta _{\mathcal{P} ,\mathcal{C}}$, $\epsilon _{\mathcal{P} ,\mathcal{C}}$, $\mu _{\mathcal{P} ,\mathcal{C}}$, $\nu _{\mathcal{P} ,\mathcal{C}}$ can be obtained using the formulas (\ref{3.2.5}), (\ref{3.2.7}), (\ref{3.2.9}), (\ref{3.2.10}), (\ref{3.2.11}), respectively. With $\upsilon $ in Eq.(\ref{3.2.6}) and $\varpi _{\mathcal{P} ,\mathcal{C}}$ in Eq.(\ref{3.2.8}) being calculated as:
\begin{equation}
\upsilon =\sum_{j=1}^{J_{\omega _k}}{\bar{w}_{\omega _k}^{\left( j \right)}\left( 1-\mathsf{p}^{\mathsf{D}}+\mathsf{p}^{\mathsf{D}}\mathsf{e}^{-\mathrm{\gamma}} \right)}
\end{equation}
\begin{equation}
\varpi _{\mathcal{P} ,\mathcal{C}}\triangleq \sum_{j=1}^{J_{\omega _k}}{\bar{w}_{\omega _k}^{\left( j \right)}\mathsf{p}^{\mathsf{D}}\frac{\mathcal{L} _{w_{k}^{\tau}}^{\left( j,\mathcal{P} ,\mathcal{C} \right)}}{\mathrm{\rho}^{\left| \mathcal{P} ,\mathcal{C} \right|}}}
\end{equation}

In addition, the calculation of the cardinality distribution $\mathscr{P} _{\pi _k}\left( \cdot \right)$ is straightforward with Eq.(\ref{3.2.4}), and the PGFs $\mathscr{G} _{\mathrm{FA}}\left( \cdot \right) $, $\mathscr{G} _{\boldsymbol{z}}\left( \cdot \middle| \boldsymbol{\xi } \right)$ and $\mathscr{G} _{k|k-1}\left( \cdot \right) $ here are calculated as:
\begin{equation}
\mathscr{G} _{\mathrm{FA}}^{\left( \left| \mathcal{P} ,\mathcal{C} \right| \right)}\left( 0 \right) =\left( \left| \mathcal{P} ,\mathcal{C} \right|! \right) \mathsf{e}^{-\mathrm{\lambda}}\,\,\mathrm{and} \,\, \mathscr{G} _{\mathrm{FA}}\left( 0 \right) =\mathsf{e}^{-\mathrm{\lambda}}
\end{equation}
\begin{equation}
\mathscr{G} _{\boldsymbol{z}}^{\left( \left| \mathcal{P} ,\mathcal{C} \right| \right)}\left( 0 \right) =\left( \left| \mathcal{P} ,\mathcal{C} \right|! \right) \mathsf{e}^{-\mathrm{\gamma}}\,\,\mathrm{and}  \,\, \mathscr{G} _{\boldsymbol{z}}\left( 0 \right) =\mathsf{e}^{-\mathrm{\gamma}}
\end{equation}
\begin{equation}
\mathscr{G} _{k|k-1}^{\left( \left| \mathcal{P} \right| \right)}\left( \upsilon \right) =\left( \left| \mathcal{P} \right|\text{!} \right) \upsilon ^{\left| \mathcal{P} \right|}\mathscr{P} _{\omega _k}\left( \left| \mathcal{P} \right| \right) 
\end{equation}

\subsection{Pruning and Merging}
Similar to the traditional PHD filter or CPHD filter, the PHD of both the GM-TPHD-E filter and the GM-TPHD-E filter has an exponentially increasing number of components over time. In order to lower the computational cost, we employ pruning and merging techniques similar to \cite{2012GranströmTSP}, that is, pruning with threshold $\mathsf{T}_{\mathsf{p}}$, setting the maximum number of components and merging components $\mathsf{J}_{\max}$ based on a given merging threshold. 

But there are some differences in our merging process: we realize that the target $\boldsymbol{\xi}$ is composed of two independent parts: the kinematic state $\boldsymbol{r}$ and the shape state $\boldsymbol{s}$. We thus split each component in PHD into two Gaussian distributions, and their merging thresholds should be specified as: kinematic merging threshold $\mathsf{T}_{\mathsf{Mr}}$ and shape merging threshold $\mathsf{T}_{\mathsf{Ms}}$, respectively. Generally, $\mathsf{T}_{\mathsf{Ms}}$ is less than $\mathsf{T}_{\mathsf{Mr}}$. 

Merging methods specific to Gaussian mixtures can be found in \cite{2012GranströmTSP}. And we give the pseudo-code of the pruning and merging for both the GM-TPHD-E filter and the GM-TCPHD-E filter in Algorithm \ref{Algorithm1}.
\begin{algorithm}[!h]
    \caption{Pruning and Merging for the GM-TPHD-E Filter and the GM-TCPHD-E Filter}
    \label{Algorithm1}
    \renewcommand{\algorithmicrequire}{\textbf{Input:}}
    \renewcommand{\algorithmicensure}{\textbf{Output:}}
    
    \begin{algorithmic}[1]
        \REQUIRE Posterior PHD parameters $\left\{ w_{k}^{\left( j \right)},t_{k}^{\left( j \right)},\boldsymbol{\xi }_{k}^{\left( j \right)},\boldsymbol{\varXi }_{k}^{\left( j \right)} \right\} _{j=1}^{J_k}$, pruning threshold $\mathsf{T}_{\mathsf{P}}$, merging threshold $\mathsf{T}_{\mathsf{Mr}}$ and $\mathsf{T}_{\mathsf{Ms}}$, maximum number of components $\mathsf{J}_\mathsf{max}$.  
        \ENSURE  Pruned posterior PHD parameters $\left\{ w_{k}^{\mathsf{o}\left( j \right)},t_{k}^{\mathsf{o}\left( j \right)},\boldsymbol{\xi }_{k}^{\mathsf{o}\left( j \right)},\boldsymbol{\varXi }_{k}^{\mathsf{o}\left( j \right)} \right\} _{j=1}^{\mathring{J}_k}$.    
        
        \STATE  Set $l=0$ and $I=\left\{ j\in \left\{ 1,2,...,J_k \right\} :w_{k}^{\left( j \right)}>\mathsf{T}_{\mathsf{P}} \right\}$.
        
        \WHILE{$I\ne \oslash $}
            \STATE $l\gets l+1$.
            \STATE $j=\underset{i\in I}{\mathrm{arg}\max} \ w_{k}^{\left( i \right)}$.
            \STATE $L_{\boldsymbol{r}}=\left\{ i\in I,\left( \boldsymbol{r}_{k}^{\left( i \right)}-\boldsymbol{r}_{k}^{\left( j \right)} \right) ^{\mathrm{T}}\left( \boldsymbol{\varXi }_{k}^{\boldsymbol{r}\left( j \right)} \right) ^{-1}\left( \cdot \right) \leqslant \mathsf{T}_{\mathsf{Mr}} \right\} $.
            \STATE $L_{\boldsymbol{s}}=\left\{ i\in I,\left( \boldsymbol{s}_{k}^{\left( i \right)}-\boldsymbol{s}_{k}^{\left( j \right)} \right) ^{\mathrm{T}}\left( \boldsymbol{\varXi }_{k}^{\boldsymbol{s}\left( j \right)} \right) ^{-1}\left( \cdot \right) \leqslant \mathsf{T}_{\mathsf{Ms}} \right\} $.
            \STATE $L=L_{\boldsymbol{r}}\cap L_{\boldsymbol{s}}$.
            \STATE Merge the corresponding components in $L$ onto the $j$-th component, refer to \cite{2012GranströmTSP}.
            \STATE $I\gets I\setminus L$.
        \ENDWHILE
        
        \IF {$l>\mathsf{J}_\mathsf{max}$}
            \STATE Only keep the $\mathsf{J}_\mathsf{max}$ components with highest weight.
        \ENDIF
        
    \end{algorithmic}
\end{algorithm}

\subsection{Estimation}
In state estimation, conventional methods can be used to extract the state of the trajectory in the posterior PHD, i.e, for the GM-TPHD-E filter, the number of trajectories is estimated as:
\begin{equation}
\hat{N}_k=\mathrm{round}\left( \sum_{j=1}^{J_k}{w_{k}^{\left( j \right)}} \right) 
\end{equation}
Then, the estimated result of trajectories can be represented as set $\left\{ \left( t_{k}^{\left( l_1 \right)},\boldsymbol{\xi }_{k}^{\left( l_1 \right)} \right) ,...,\left( t_{k}^{\left( l_{\hat{N}_k} \right)},\boldsymbol{\xi }_{k}^{\left( l_{\hat{N}_k} \right)} \right) \right\}$ where $\left\{ l_1,...,l_{\hat{N}_k} \right\}$ are the indices of the PHD components with highest weights.

For the GM-TCPHD-E filter, the number of trajectories can be obtained as:
\begin{equation}
\hat{N}_k=\underset{n\in \mathbb{N} \cup \left\{ 0 \right\}}{\mathrm{arg}\max}\mathscr{P} _{\pi _k}\left( n \right) 
\end{equation}

Then the state estimation process is consistent with the GM-TPHD-E filter.

\subsection{Discussion}
In this section, we discuss the computational complexity and scalability aspects of the proposed filters. As mentioned above, the main difference between the TST-based PHD/CPHD filters and their ordinary implementation is that the TST-based filters do not integrate out previous states of the trajectories \cite{2019GarcíaTSP}. This means that in filtering recursion, the introduction of trajectory sets is only for storing the target state at each time instant and does not cause an increase in the components of PHD, so our proposed filters have no significant difference in computational complexity compared to previous works, such as \cite{2012GranströmTSP,2013LundquistJSTSP}. However, our methods also introduce additional shape parameter estimation, which performs sequential measurement updates, resulting in slightly higher overall time costs than these previous filters.

We also would like to note that to simplify the exposition, the linear Gaussian state space model is considered, while the detection probability $\mathsf{p}^{\mathsf{D}}$ and the measurement rate $\mathrm{\gamma}$ are assumed to be constants. However, it is evident that our method can be applied to nonlinear situations by combining extended Kalman filtering or other iterated posterior linearization filters \cite{Koch2014TrackingAS}. For time-varying detection probability and target measurement rate, they can be modeled as state independent random variables and estimated using Beta distribution and Gamma distribution, respectively. Methods of iteratively estimating these distribution parameters are presented in, e.g., \cite{2021WeiICSIP,2022WeiTVT2}. The presented filters can easily be augmented in a similar manner to provide these parameters over time.

\section{PERFORMANCE EVALUATION}
\label{Sec6}
In this section, we compare the performance of the proposed methods-the GM-TPHD-E filter and the GM-TCPHD-E filter in METT with existing tracking methods and focus on their accuracy in estimating the extended shape of targets and generating target trajectories.

\subsection{Contrast Methods}
\label{Sec6-A}
The methods selected for comparison include the following.
\begin{itemize}
    \item The GM-TPHD-GIW filter: In this method, we use TPHD combined with standard RMM to achieve tracking and shape estimation of multiple targets, and implement it in the form of the Gaussian mixture. The details of the implementation can be found in \cite{2021SjudinFUSION}. 
    \item The GM-LPHD-E filter: In this method, we use the labeling approach \cite{2009PantaTAES} to establish trajectories for each component of the Gaussian mixture PHD filter, and update the extended shape of the multiple targets in the same way as the proposed method.
\end{itemize}

Note that among the two METT methods mentioned above, the former focuses on comparing the estimation of the extended shape of targets, while the latter tends to compare the generation of the trajectories. Therefore, their introduction is reasonable and necessary.

\subsection{Experimental Scenarios}
The 2-dimensional Cartesian coordinate is used to define the measurement and the state parameters of the target, i.e., $d=2$. We tested our methods in two scenarios, the first being the simulation scenario we constructed, and the second being generated based on traffic monitoring data from a signalized intersections in the real world. In these scenarios, the time step interval $\mathrm{T}=1s$, and all targets are assumed to follow a uniform and time-invariant linear Gaussian constant velocity model. The kinematic state of a single target is considered as $\boldsymbol{r}=\left[ o_x,o_y,\dot{o}_x,\dot{o}_y \right] ^{\mathrm{T}}$, which includes the position and velocity shown in Eq.(\ref{4-33}). The shape state of a single target is described as an elliptical, i.e, $\boldsymbol{s}=\left[ \theta ,l_1,l_2 \right] ^{\mathrm{T}}$, presented in Eq.(\ref{4-34}).

The dynamic matrix of the targets' kinematic state and the corresponding process noise matrix are set as:
\begin{equation}
\mathsf{F}^{\boldsymbol{r}}=\left[ \begin{matrix}
	1&		\mathrm{T}\\
	0&		1\\
\end{matrix} \right] \otimes \mathbf{I}_2\,\,\mathrm{and} \,\,\mathsf{Q}^{\boldsymbol{r}}=\left( \mathsf{q}^{\boldsymbol{r}} \right) ^2\cdot \left( \left[ \begin{matrix}
	{{\mathrm{T}^3}/{3}}&		{{\mathrm{T}^2}/{2}}\\
	{{\mathrm{T}^2}/{2}}&		\mathrm{T}\\
\end{matrix} \right] \otimes \mathbf{I}_2 \right) 
\end{equation}

The dynamic matrix of the targets' shape state and the corresponding process noise matrix are set as:
\begin{equation}
\mathsf{F}^{\boldsymbol{s}}=\mathbf{I}_3\,\,\mathrm{and}  \,\,\mathsf{Q}^{\boldsymbol{s}}=\mathrm{diag}\left( \left( \mathsf{q}^{\theta} \right) ^2,\left( \mathsf{q}^l \right) ^2,\left( \mathsf{q}^l \right) ^2 \right) 
\end{equation}
\noindent where $\mathsf{q}^{\boldsymbol{r}}$ is the standard deviation of the kinematic noise, $\mathsf{q}^{\theta}$ and $\mathsf{q}^l$ are the standard deviations of the orientation and the length of the axis, respectively.

The parameters of the measurement model are given as:
\begin{equation}
\mathsf{H}=\left[ \left[ \begin{matrix}
	1&		0\\
\end{matrix} \right] \otimes \mathbf{I}_2,\mathbf{0}_{2\times 3} \right] \,\, \mathrm{and} \,\, \mathsf{Q}^{\boldsymbol{e}}=\left( \mathsf{q}^{\boldsymbol{e}} \right) ^2\cdot \mathbf{I}_2
\end{equation}
\noindent where $\mathsf{q}^{\boldsymbol{e}}$ is the standard deviation of the measurement noise.

\subsection{Metrics}
In order to evaluate the accuracy of the shape estimation of targets with an elliptical extent, the Gaussian Wasserstein Distance (GWD) metric proposed in \cite{2016YangIEEEConfSyst} has been adopted. Given the kinematic states of two elliptical targets are $\boldsymbol{r}_1$ and $\boldsymbol{r}_2$, with corresponding extended matrices $\boldsymbol{X}_1$ and $\boldsymbol{X}_2$, then the GWD between them can be calculated as:
\begin{equation}
\label{GWD_metric}
d_{\mathsf{GW}}=\left\| \boldsymbol{r}_1-\boldsymbol{r}_2 \right\| ^2+\mathrm{tr}\left( \boldsymbol{X}_1+\boldsymbol{X}_2-2\sqrt{\sqrt{\boldsymbol{X}_1}\boldsymbol{X}_2\sqrt{\boldsymbol{X}_1}} \right) 
\end{equation}

\noindent where $\mathrm{tr}\left( \cdot \right)$ represents the trace of the matrix, and these two ellipses can be regarded as the true and estimated values of the target’s state. Note that in our modeling, the extended matrix $\boldsymbol{X}$ of the target can be calculated using its shape parameters $\boldsymbol{s}=\left[ \theta ,l_1,l_2 \right] ^{\mathrm{T}}$:
\begin{equation}
\boldsymbol{X}=\boldsymbol{SS}^{\mathrm{T}}
\end{equation}
\begin{equation}
\boldsymbol{S}=\left[ \begin{matrix}
	\cos \theta&		-\sin \theta\\
	\sin \theta&		\cos \theta\\
\end{matrix} \right] \left[ \begin{matrix}
	l_1&		0\\
	0&		l_2\\
\end{matrix} \right] 
\end{equation}

In addition, since all methods are based on RFS, we further introduced the Trajectory Metric (TM) \cite{2020GarcíaTSP} based on the GWD \footnote{ The encapsulated TM code can be found in: https://github.com/yuhsuansia. Thanks to the authors for their excellent work.} to judge the ability to correctly generate target trajectories. The TM consists of four parts, i.e, the location cost, the missed cost, the false cost, and the track switch cost, and these four parts are considered comprehensively to obtain the overall TM error. The TM parameters are set as: location/extent error cut-off $c$=40, order $p$=1, and track switch cost $a$=2. 

We evaluated the performance of the filters using Monte Carlo simulation (MC) with 100 runs\footnote{The MATLAB implementations of the filters have been run on a workstation with a 2.8 GHz Intel Core i9-10900 CPU}. All metrics were averaged over 100 MCs, and we recorded the average time consumption of each method in single MC as the measure of computational efficiency. All quantity units in this section are based on the international system of units.

\subsection{Scenario 1: the Simulation Scene}
\label{Sec6-D}
\begin{figure}[!htp]
    \centering
    \includegraphics[width=3in]{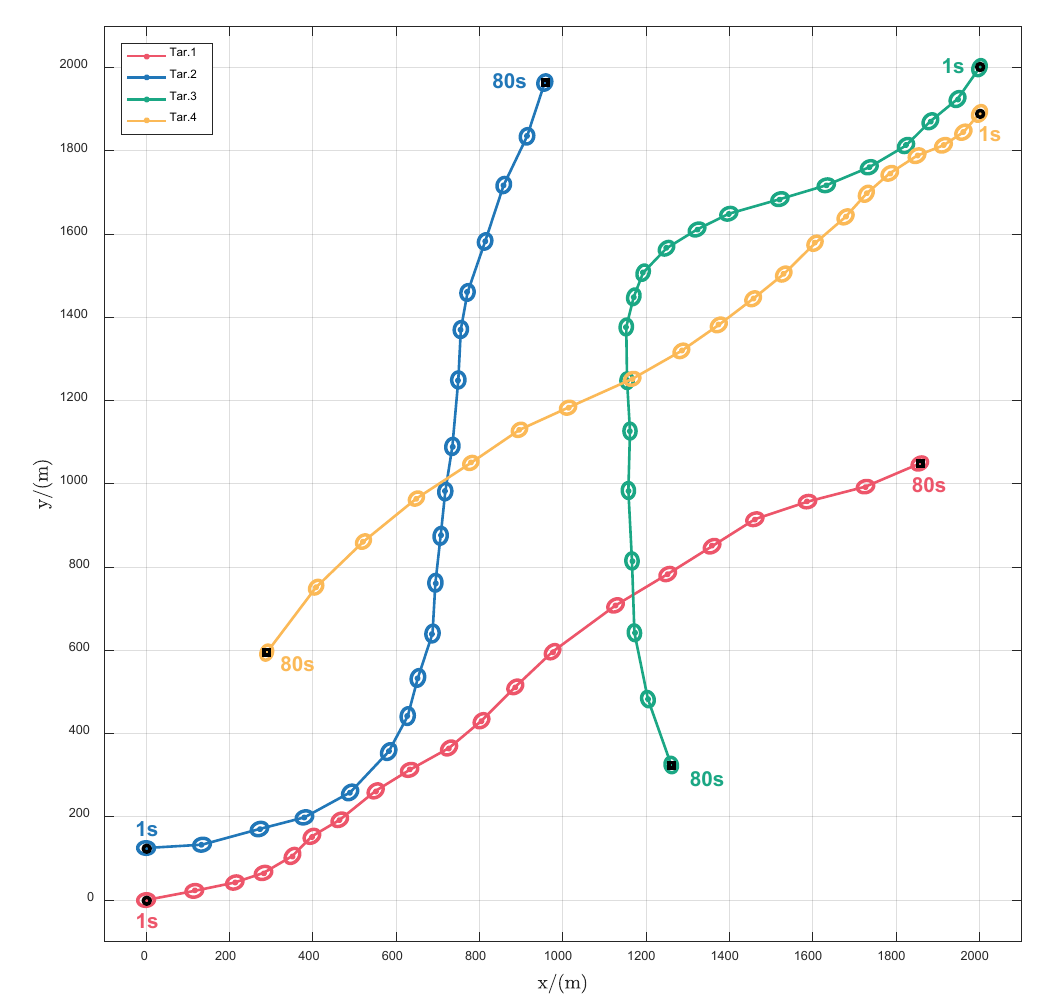}
    \caption{The Ground Truth (GT) of the Scenario 1. \\
    (\small{We draw the position of the target every four seconds, and for each target trajectory, we use `$\circ$' to denote the starting point and `$\square$' to indicate the ending point})}
    \label{fig:f1-1}
\end{figure}

In this simulation scenario, we consider the motion of the targets illustrated in Fig.\ref{fig:f1-1}, including 4 targets simulated over 80 time steps. The size of the scene is [-100,2100]$\times$[-100,2100]. At the initial time step of the simulation, the positions of the four targets are: Target 1(Tar.1, the red one):[0,0]; Tar.2(the blue one):[0,125]; Tar.3(the green one):[2000,2000]; Tar.4 (the yellow one):[2000,1890]. After starting the simulation, all four targets move towards the center of the scene, and the axis length of their elliptical extents remains as [40,30] during the motion. It should be noted that at 15s$\sim$30s, Tar.3 and Tar.4 underwent parallel motion in the same direction within close range, and the number of targets does not change throughout the simulation process.

The filtering parameters of all methods are consistently, set as:
\begin{itemize}
    \item The standard deviation of process noise: $\mathsf{q}^{\boldsymbol{r}}=10, \mathsf{q}^{\theta}=0.05$ and $\mathsf{q}^l=0.1$.
    \item The standard deviation of measurement noise: $\mathsf{q}^{\boldsymbol{e}}=10$.
    \item The detection probability: $\mathsf{P}^{\mathsf{D}}=0.98$.
    \item The survival probability: $\mathsf{P}^{\mathsf{S}}=0.99$.
    \item The target measurement rate: $\mathsf{\gamma} =20$.
    \item Uniformly distributed Poisson clutter with rate $\lambda =10$.
    \item The birth density consists of four components with weight $w_{k}^{\beta ,\left( j \right)}=0.1, \forall j,k$, their positions are determined by adding a deviation $\backepsilon \sim \mathcal{N} \left( 0,\left[ 10,10 \right] \right)$ to the true initial positions of each target. Their velocities are all $\left[ 0,0 \right] ^{\mathrm{T}}$, and their shape states are all $\left[ 0,45,35 \right] ^{\mathrm{T}}$ \footnote{Compared to the size of the targets’ actual shape, this setting is relatively large}. In addition, each component has the initial kinematic state covariance $\mathrm{diag}\left[50,50,5,5\right]$, and the shape state covariance $\mathrm{diag}\left[ 0.2,100,100 \right]$.
    \item The pruning thresholds are established as: the pruning threshold $\mathsf{T}_{\mathsf{p}}$ = 1e-5,  the maximum number of components $\mathsf{J}_{\mathsf{\max}}$ = 300. For the proposed filters, the kinematic merge threshold $\mathsf{T}_{\mathsf{Mr}}$ = 4, the shape merge threshold $\mathsf{T}_{\mathsf{Ms}}$ = 1. For the GM-TPHD-GIW filter, the Gaussian merging threshold $\mathsf{T}_{\mathsf{N}}$ = 4, the inverse Wishart merging threshold $\mathsf{T}_{\mathsf{IW}}$ = 3.
\end{itemize}

\begin{figure*}[!htp]
    \centering
    \includegraphics[width=7in]{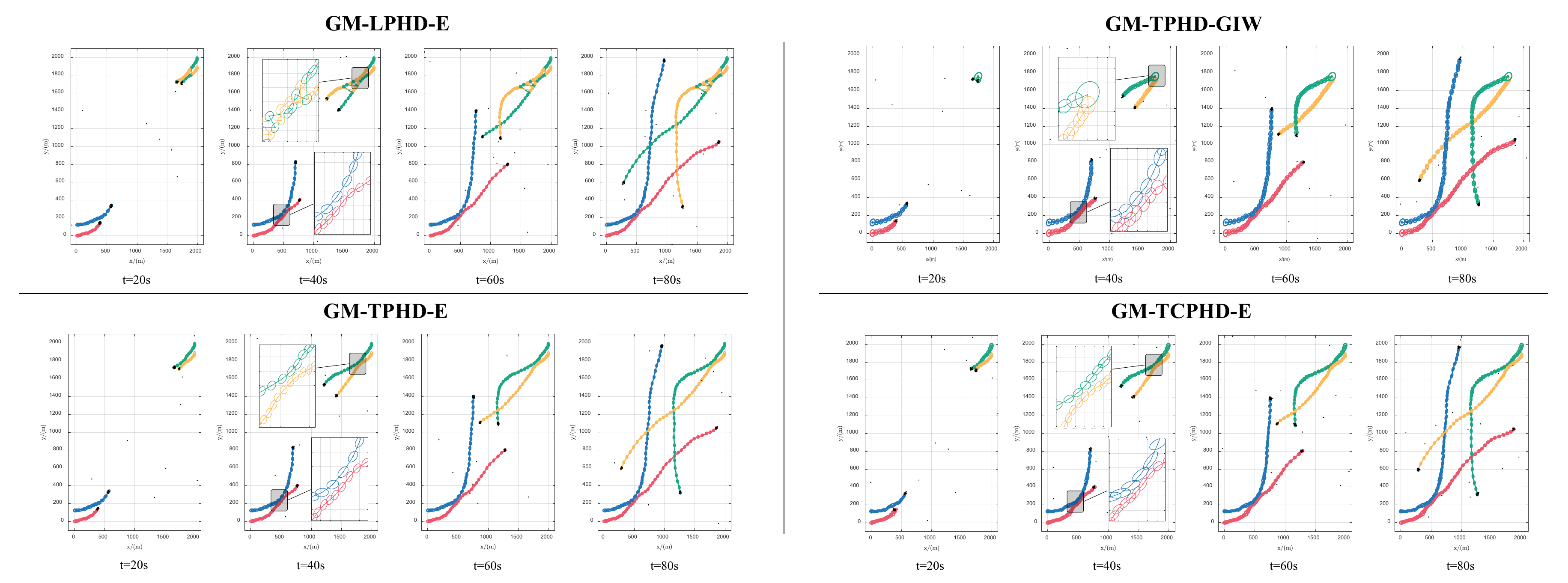}
    \caption{One sample simulation of four methods for Scenario 1. \\(\small{We have selected several representative time steps for presentation, including the $\left\{20, 40, 60, 80\right\}$ second. Different colors are used to distinguish different target trajectories, and we use black dots to represent the measurements at each time step. The GM-TPHD-E filter and the GM-TCPHD-E filter are able to provide the complete and accurate trajectories of the targets at each time step. As a comparison, the GM-TPHD-GIW filter and the GM-LPHD-E filter do not work well, the former has shortcomings in the initiation of the trajectories and shape estimation of the target, while the latter exhibits a phenomenon of 'trajectory switching' in generating trajectories.})}
    \label{fig:f1-2}
\end{figure*}

An sample set of outputs of the four methods is shown in Fig.\ref{fig:f1-2}. It can be seen that the GM-TPHD-GIW filer, based on the RMM, cannot capture changes in the axis length of elliptical targets. Given a larger initial shape, the shape estimation of the targets will also be larger, making this method difficult to accurately estimate the shape parameters of the targets and accompanied by a certain degree of trajectory start lag phenomenon (see the 40s state of this method). The GM-LPHD-E filter shows considerably worse performance, as two targets are born at the same time step and have a similar close range motion; this method confuses the trajectories of the two targets and causes the phenomenon of ‘trajectory switch’. In fact, the method of generating trajectories through labels can also cause such two targets to share the same trajectory, resulting in trajectory loss \cite{2019GarcíaTSP}. It seems that the two methods, the GM-TPHD-E filter and the GM-TCPHD-E filter, provide accurate target trajectory information at each time step, and along with the tracking process, the shape estimation of the target also converges better to its true value, a more precise ellipse.

\begin{figure*}[!t]
    \centering
    \includegraphics[width=6.5in]{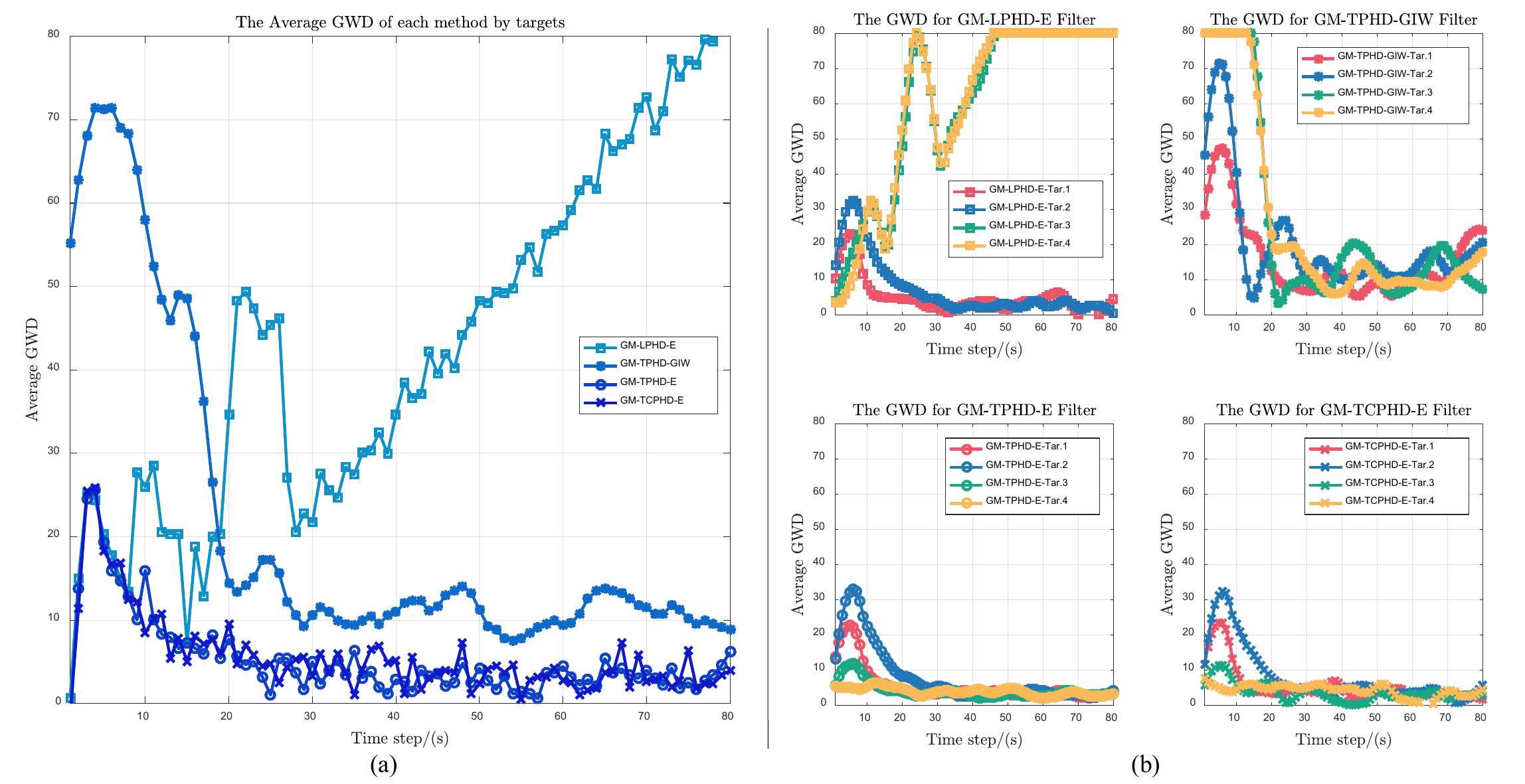}
    \caption{The GWD result for Scenario 1.}
    \label{fig:f1-3}
\end{figure*}

The GWDs of the four methods averaged over the 100 MCs are plotted in Fig.\ref{fig:f1-3}(b), and we further average the GWDs of the four targets and present the results in Fig.\ref{fig:f1-3}(a). As expected, the GWDs of the two proposed methods eventually converge and stabilize within an acceptable range. For the GM-TPHD-GIW filter, due to the delayed initialization of the trajectory, its GWD between 1s$\sim$10s is very large and gradually converges in subsequent tracking. However, due to the inaccuracy of its shape estimation, the convergence value is significantly higher than that of the proposed methods. For the GM-LPHD-E filter, it has a relatively high error. Due to confusion of the trajectories between Tar.3 and Tar.4, their GWD peaks appear within 15s$\sim$30s, and in the subsequent process, the GWDs of the two targets gradually diverges due to the trajectories switching in the subsequent process.

\begin{figure}[H]
    \centering
    \includegraphics[width=3.5in]{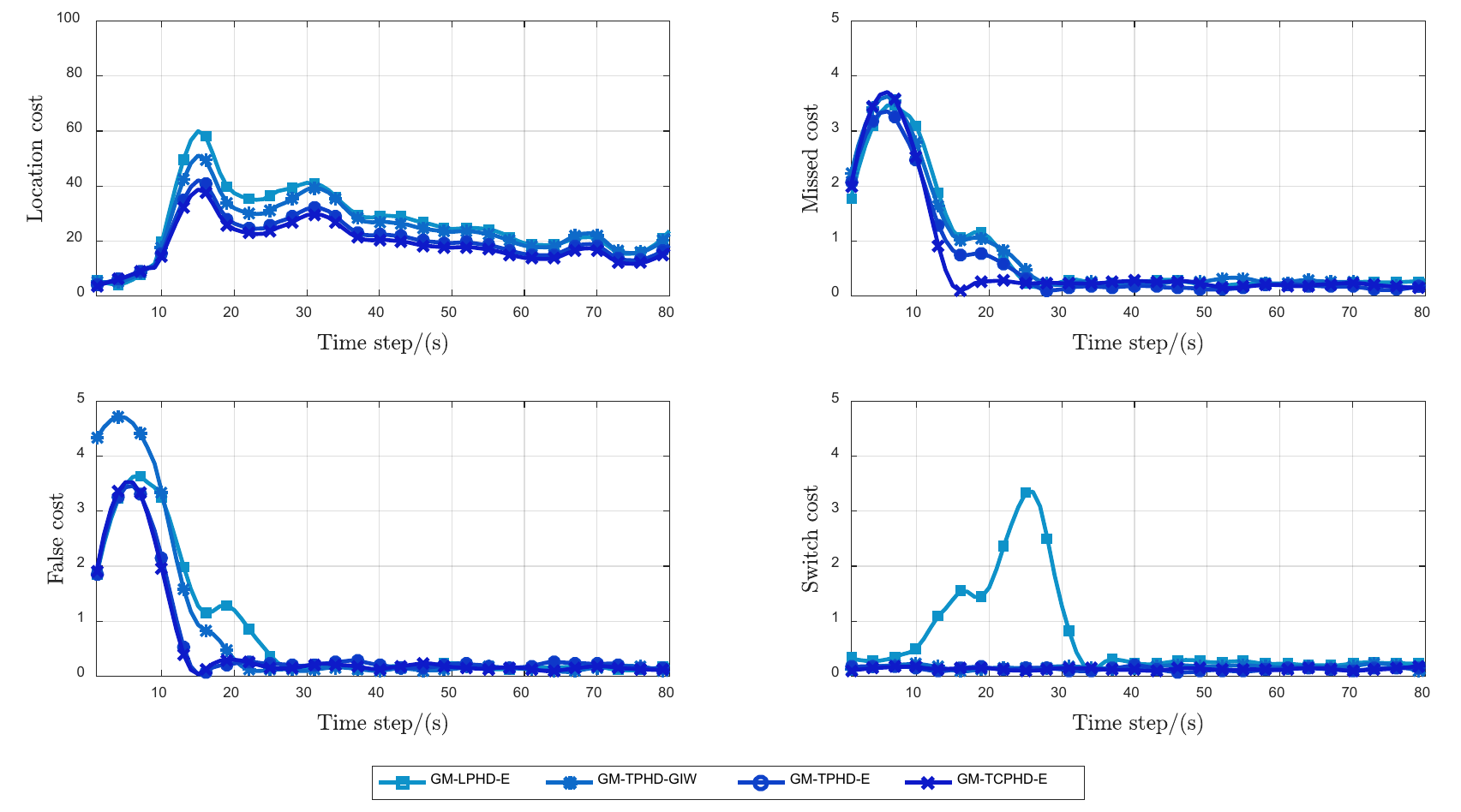}
    \caption{The TM result for Scenario 1.}
    \label{fig:f1-4}
\end{figure}

In addition, to analyze the results more thoroughly, we present the average TM results of the four methods in Fig.\ref{fig:f1-4} to evaluate the precision of all methods in generating target trajectories. It can be seen that the GM-TPHD-GIW filter has the highest location cost because of its disadvantage in shape estimation. The GM-LPHD-E filter has a higher switch cost before the 30s, which is the result of trajectory confusion. In addition, we can also see that the two proposed methods perform very similarly overall. However, the GM-TCPHD-E filter is slightly better than the GM-TPHD-E filter in term of miss cost, which is derived from the advantages of the CPHD-based recursion in the number of targets estimation.

The running times of the MATLAB implementations of the filters are displayed in Tab.\ref{Tab.f1-1}. It can be seen that, due to the introduction of more complex shape estimation, the proposed method is slower than the GM-TPHD-GIW filter. On the other hand, due to the advantages of TST in track generation calculations, our methods have less time cost than the labeling-based GM-LPHD-E filter.

\begin{table}[!h]
\caption{RUNNING TIMES OF THE FILTERS}
\label{Tab.f1-1}
\centering
\begin{tabular}{|c|cc|cc|}
\hline
\multicolumn{1}{|l|}{}   & \multicolumn{2}{c|}{Scenario 1}            & \multicolumn{2}{c|}{Scenario 2}            \\ \hline
\multirow{2}{*}{Filters} & \multicolumn{2}{c|}{Time cost}             & \multicolumn{2}{c|}{Time cost}             \\ \cline{2-5} 
                         & \multicolumn{1}{c|}{each MC} & cycle times & \multicolumn{1}{c|}{each MC} & cycle times \\ \hline
GM-LPHD-E                & \multicolumn{1}{c|}{69.97}   & 0.87        & \multicolumn{1}{c|}{112.12}  & 1.24        \\ \hline
GM-TPHD-GIW              & \multicolumn{1}{c|}{52.27}   & 0.65        & \multicolumn{1}{c|}{91.16}   & 1.01        \\ \hline
GM-TPHD-E                & \multicolumn{1}{c|}{65.89}   & 0.82        & \multicolumn{1}{c|}{110.49}  & 1.23        \\ \hline
GM-TCPHD-E               & \multicolumn{1}{c|}{152.31}  & 1.72        & \multicolumn{1}{c|}{207.68}  & 2.31        \\ \hline
\end{tabular}
\end{table}

Moreover, we further show the overall TM error for other simulation parameters on the left side of Tab.\ref{Tab.f1-2}. As can be seen, the performance of all the methods was affected under different parameter settings. However, the proposed GM-CPHD-E filter achieved the best performance, which we have highlighted in bold on Tab.\ref{Tab.f1-2}. The performance of the proposed GM-PHD-E filter is slightly inferior.

\subsection{Scenario 2: the Scene with Real Data}
\label{Sec 6-E}
\begin{figure}[H]
    \centering 
        \subfigure[A real signalized intersection located in Chongqing]{%
			\resizebox*{6.5cm}{!}{\includegraphics{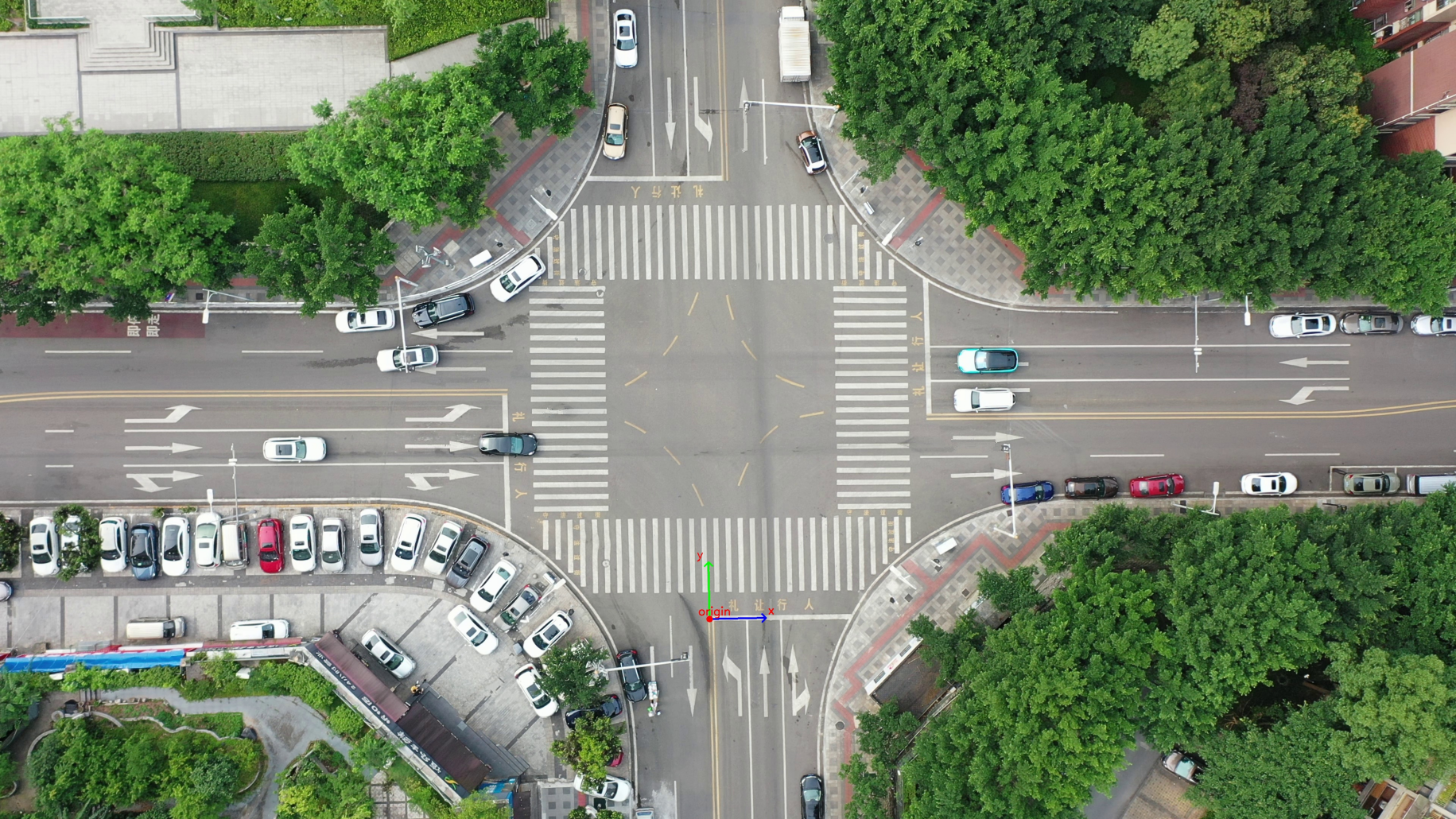}}}\hspace{5pt}
		
        \subfigure[Four entrances in Scenario 2]{%
			\resizebox*{6.5cm}{!}{\includegraphics{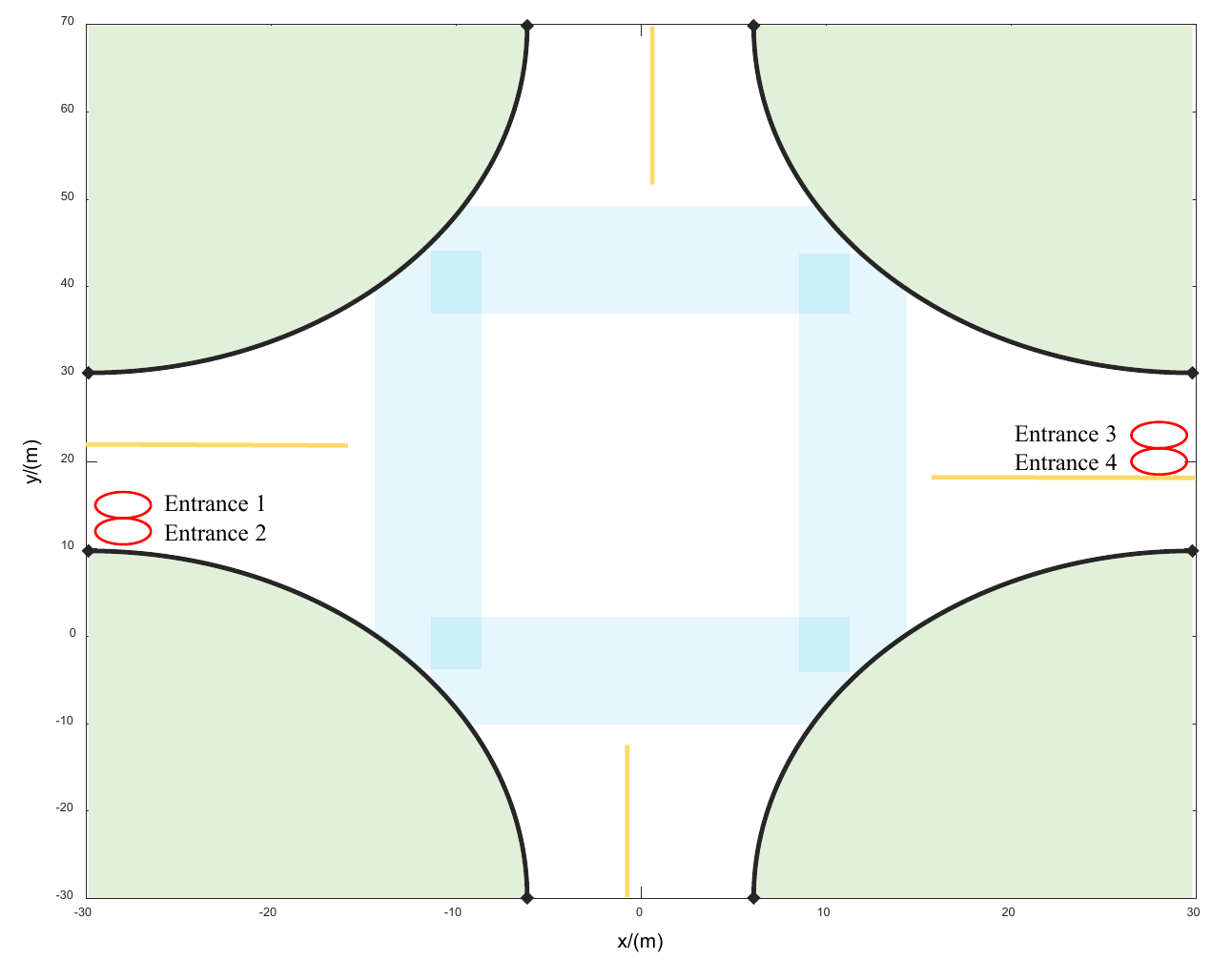}}}
		
        \subfigure[The GT of the Scenario 2]{%
			\resizebox*{6.5cm}{!}{\includegraphics{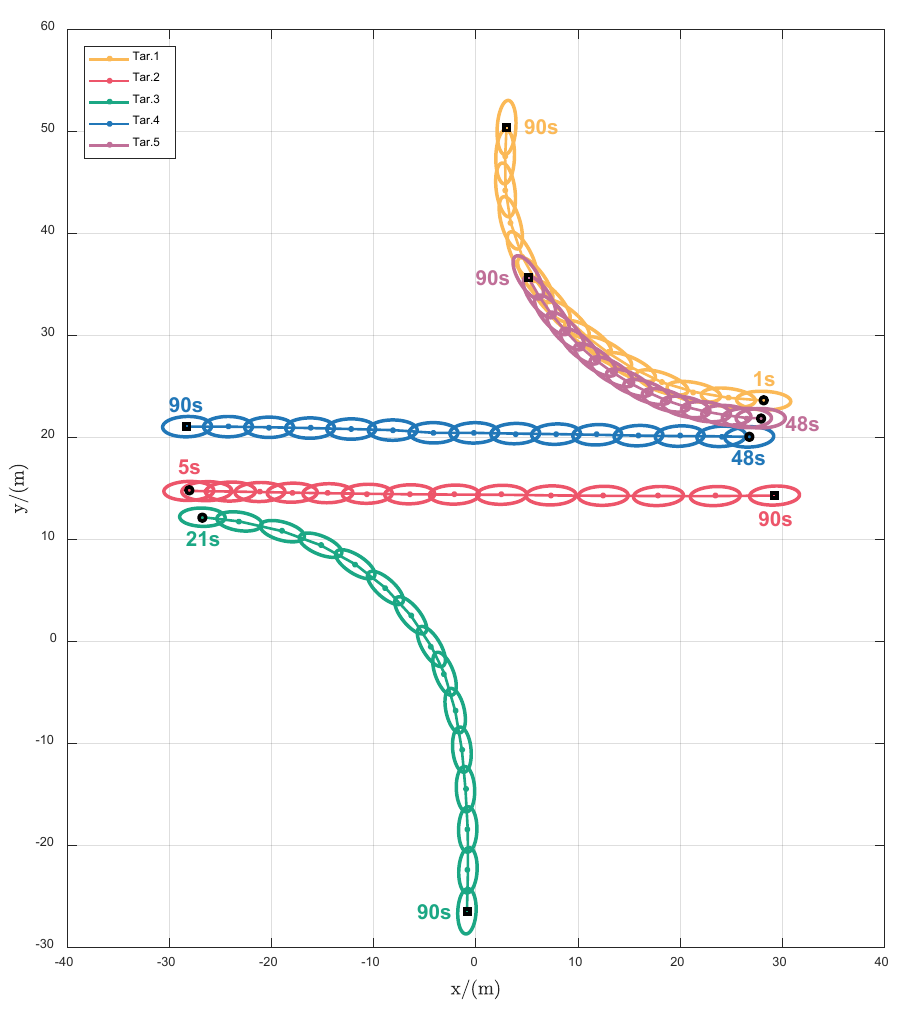}}}
    \caption{The schematic diagram of Scenario 2.}
    \label{Fig.f2-1}
\end{figure}

In this experiment, we used real data to create scenarios and further demonstrated the ability of our methods. The test data are sourced from publicly available datasets called SIND \cite{2022XuITSC}, collected in an urban signalized intersection in Chongqing with more than 13000 traffic participants. As shown in Fig. \ref{Fig.f2-1}(a), this data set uses drone aerial photography to monitor traffic and uploads images of the surveillance region every second. After that, image detection methods are used to export the position and appearance information of each vehicle. On this basis, we consider each vehicle as an extended target and refer to \cite{2021TuncerTSP} to generate measurements for each vehicle using feature extraction algorithms. And image filtering algorithms are adopted to reduce the number of clutters.

We captured a continuous period of traffic to construct the scene with size [-40,40]$\times$[-30,60], which includes a total of 5 vehicles. Due to the different entry times of each vehicle into the intersection, the number of targets in the scene is varying over time. Since all vehicles in this scene come from the left or right sides of the intersection, they will enter the scene through the four entrances shown in Fig. \ref{Fig.f2-1}(b). The dynamic process of each target is shown in Fig.\ref{Fig.f2-1}(c), which can be described as: At 1s, Tar.1 entered the scene from entrance 3 and makes a right turn. Subsequently, at 5s, Tar.2 entered from entrance 1 and went straight ahead. Then Tar.3 entered the scene from entrance 2 at the 21s and turned right after. Tar.4 and Tar.5 both appeared in the 48s and entered the scene through entrance 4 and entrance 3, respectively. Within the following 5 seconds, they underwent a very close distance, same direction and speed motion, before separating Tar.4 continued straight and Tar.5 turned right, respectively. The entire scene lasted 90 seconds and all targets are present in the scene throughout the simulation time.

More details of Scenario 2 are listed in the Tab.\ref{Tab.T2-1}. Due to the smaller scene size, the following modifications have been made to the filtering parameter settings compared to Scenario 1: 

\begin{table*}[!h]
\caption{DETAILS ABOUT THE SCENARIO 2}
\centering
\label{Tab.T2-1}
\begin{tabular}{|cccccc|}
\hline
\multicolumn{6}{|c|}{Details of the Targets}                                                                               \\ \hline
\multicolumn{1}{|c|}{Label}          & \multicolumn{1}{c|}{Size}                & \multicolumn{1}{c|}{Type} & \multicolumn{1}{c|}{Entrance}       & \multicolumn{1}{c|}{Move Time}   & Dynamic Process \\ \hline
\multicolumn{1}{|c|}{Tar.1}          & \multicolumn{1}{c|}{{[}2.6536×0.9128{]}} & \multicolumn{1}{c|}{Car}  & \multicolumn{1}{c|}{Entrance 3}     & \multicolumn{1}{c|}{{[}1,90{]}}  & Turn right  \\ \hline
\multicolumn{1}{|c|}{Tar.2}          & \multicolumn{1}{c|}{{[}2.4417×0.9318{]}} & \multicolumn{1}{c|}{Car}  & \multicolumn{1}{c|}{Entrance 1}     & \multicolumn{1}{c|}{{[}5,90{]}} & Straight ahead  \\ \hline
\multicolumn{1}{|c|}{Tar.3}          & \multicolumn{1}{c|}{{[}2.2238×0.8963{]}} & \multicolumn{1}{c|}{Car}  & \multicolumn{1}{c|}{Entrance 2}     & \multicolumn{1}{c|}{{[}21,90{]}} & Turn right      \\ \hline
\multicolumn{1}{|c|}{Tar.4}          & \multicolumn{1}{c|}{{[}2.3195×0.9961{]}} & \multicolumn{1}{c|}{Car}  & \multicolumn{1}{c|}{Entrance 4}     & \multicolumn{1}{c|}{{[}48,90{]}}  & Straight ahead      \\ \hline
\multicolumn{1}{|c|}{Tar.5}          & \multicolumn{1}{c|}{{[}2.4052×0.9546{]}} & \multicolumn{1}{c|}{Car}  & \multicolumn{1}{c|}{Entrance 3}     & \multicolumn{1}{c|}{{[}48,90{]}} & Turn right      \\ \hline
\multicolumn{6}{|c|}{Details of the Scene}                                                                                                                                                             \\ \hline
\multicolumn{1}{|c|}{Entrance Label} & \multicolumn{2}{c|}{Entrance Location}                               & \multicolumn{1}{c|}{Entrance Label} & \multicolumn{2}{c|}{Entrance Location}             \\ \hline
\multicolumn{1}{|c|}{Entrance 1}     & \multicolumn{2}{c|}{{[}-28,15{]}}                                    & \multicolumn{1}{c|}{Entrance 2}     & \multicolumn{2}{c|}{{[}-28,12{]}}                  \\ \hline
\multicolumn{1}{|c|}{Entrance 4}     & \multicolumn{2}{c|}{{[}28,23{]}}                                     & \multicolumn{1}{c|}{Entrance 4}     & \multicolumn{2}{c|}{{[}28,20{]}}                   \\ \hline
\end{tabular}
\end{table*}

\begin{itemize}
    \item The standard deviation of process noise: $\mathsf{q}^{\boldsymbol{r}}=0.5, \mathsf{q}^{\theta}=0.01$ and $\mathsf{q}^l=0.05$.
    \item The standard deviation of measurement noise: $\mathsf{q}^{\boldsymbol{e}}=1$.
    \item The target measurement rate: $\mathsf{\gamma} =10$.
    \item Uniformly distributed Poisson clutter with rate $\lambda =20$. This means that Scenario 2 has more dense clutter.
    \item The birth density consists of four components corresponding to the four entrances of the scene, respectively, and their parameter settings are similar to Scenario 1, except that their initial shape parameters are set to $[0,1.5,1.5]^{\mathrm{T}}$. The initial kinematic state covariance and shape state covariance are $\mathrm{diag}\left[ 0.5,0.5,0.1,0.1 \right] $ and $\mathrm{diag}\left[ 0.01,0.5,0.5 \right]$, respectively.
    \item The thresholds related to pruning have been reduced tenfold.
\end{itemize}

\begin{figure*}[!htp]
    \centering
    \includegraphics[width=7in]{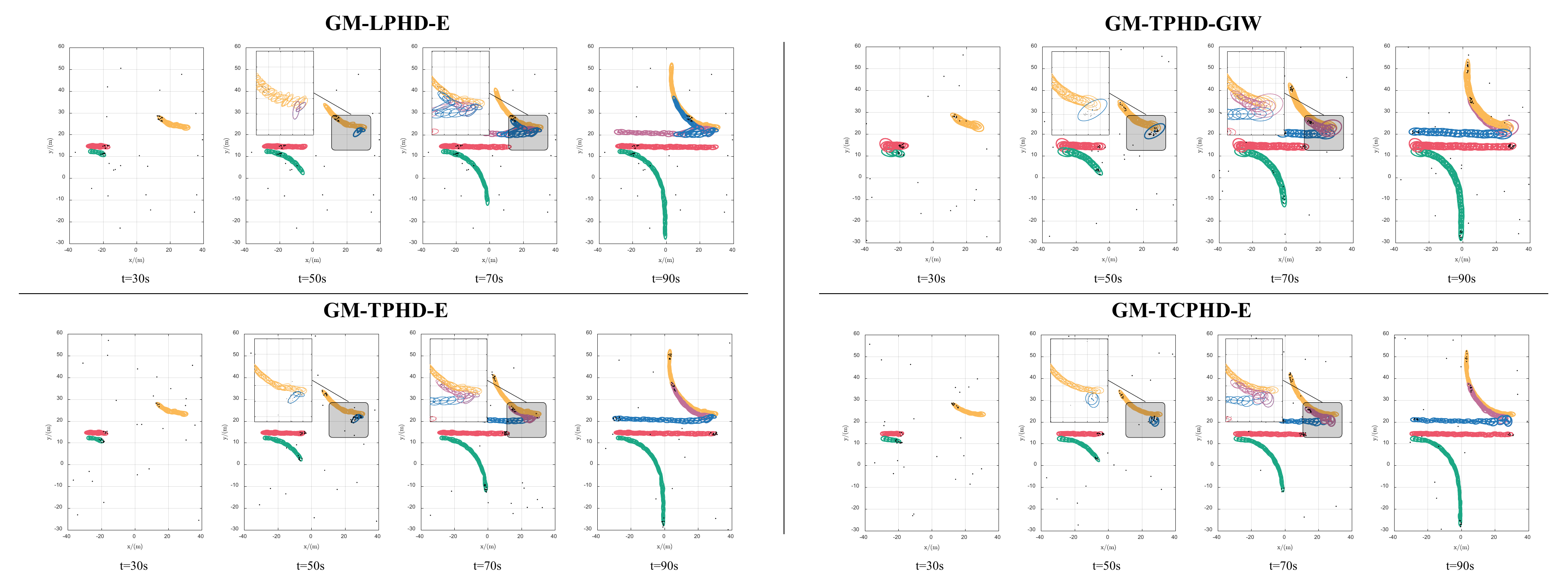}
    \caption{One example simulation of four methods for Scenario 2.}
    \label{fig:f2-*}
\end{figure*}

\begin{figure*}[!t]
    \centering
    \includegraphics[width=6.5in]{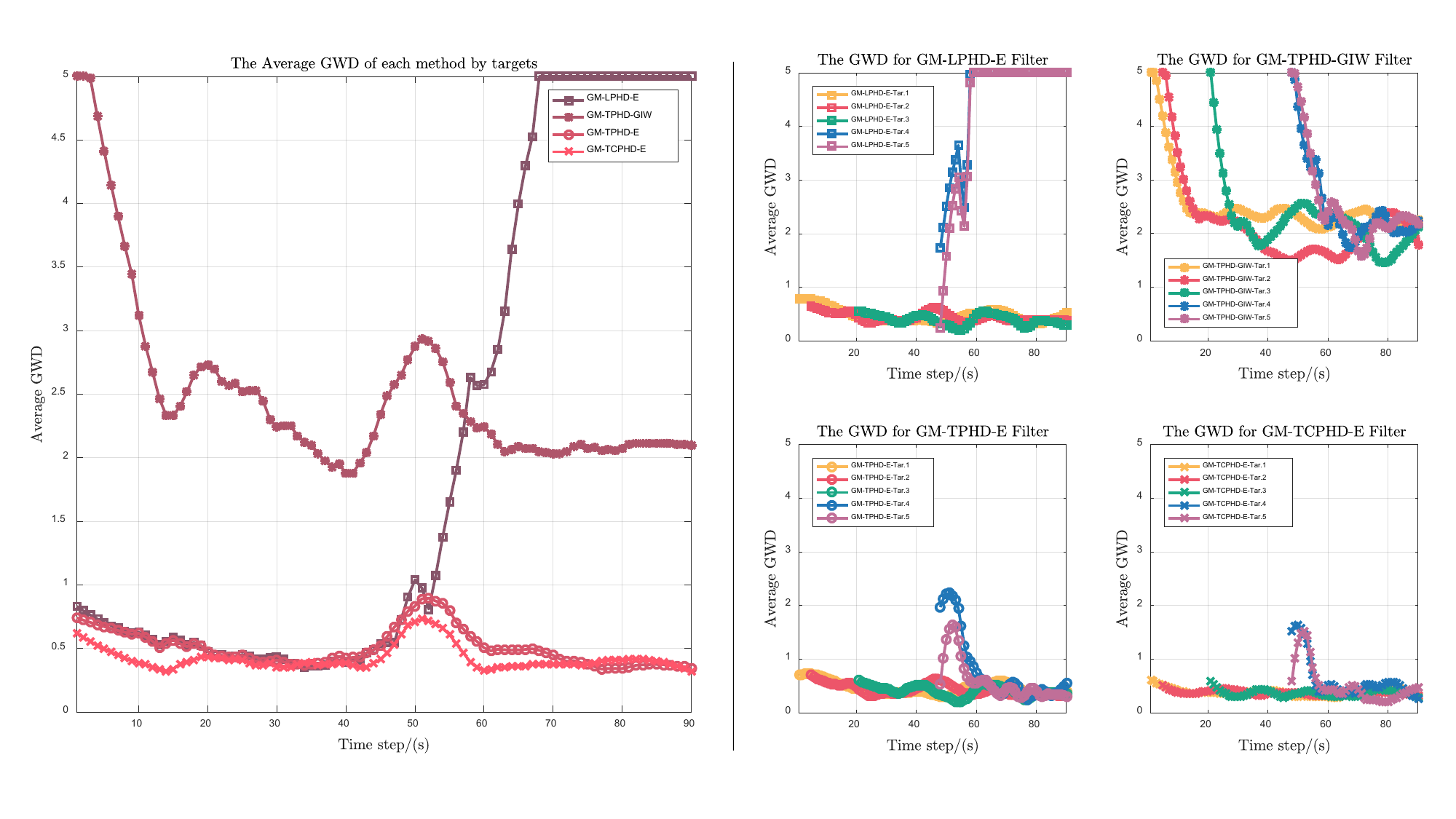}
    \caption{The GWD result for Scenario 2.}
    \label{fig.f2-3}
\end{figure*}

\begin{figure}[H]
    \centering
    \includegraphics[width=3.5in]{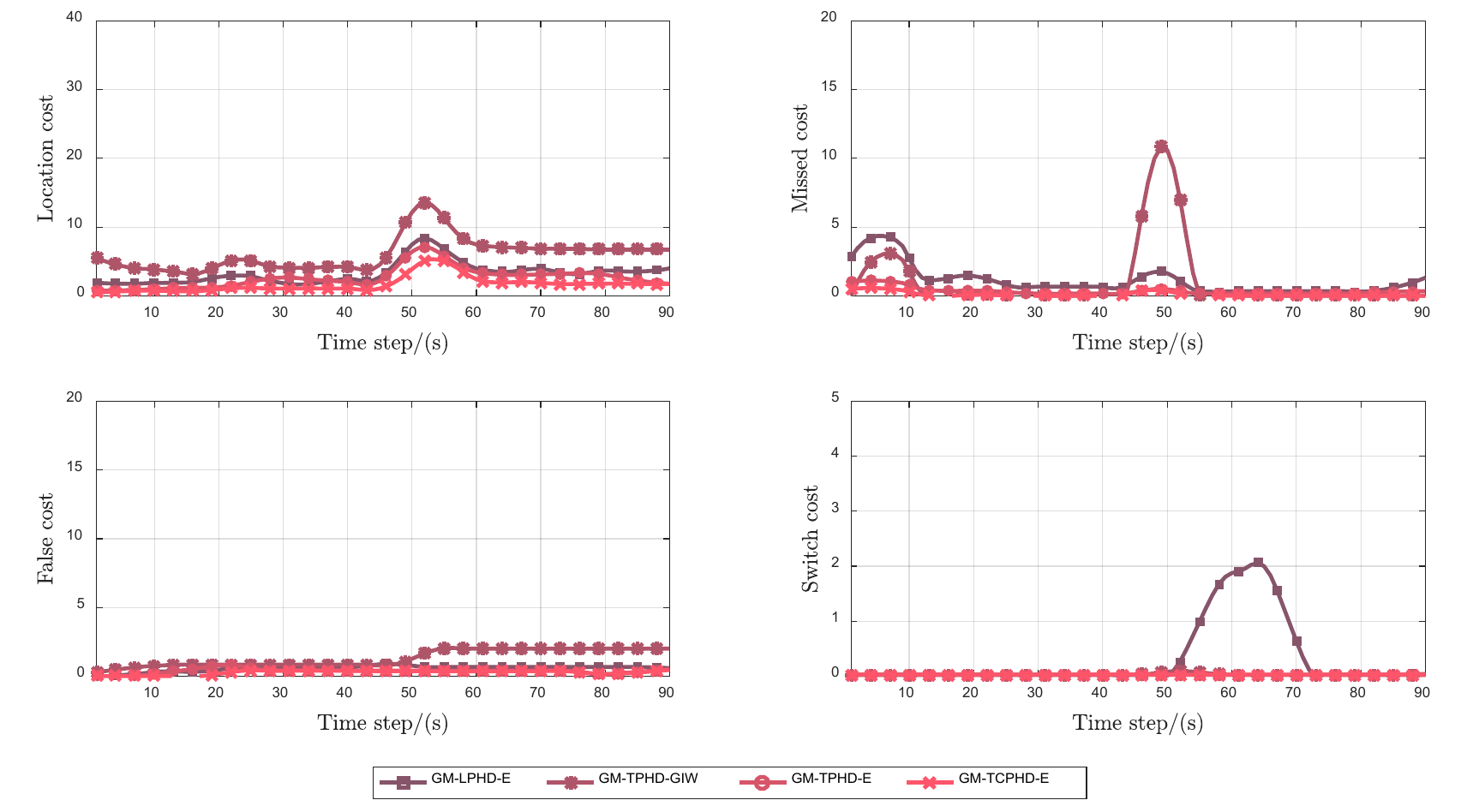}
    \caption{The TM result for Scenario 2.}
    \label{fig.f2-4}
\end{figure}

\begin{figure}[H]
    \centering
    \includegraphics[width=3in]{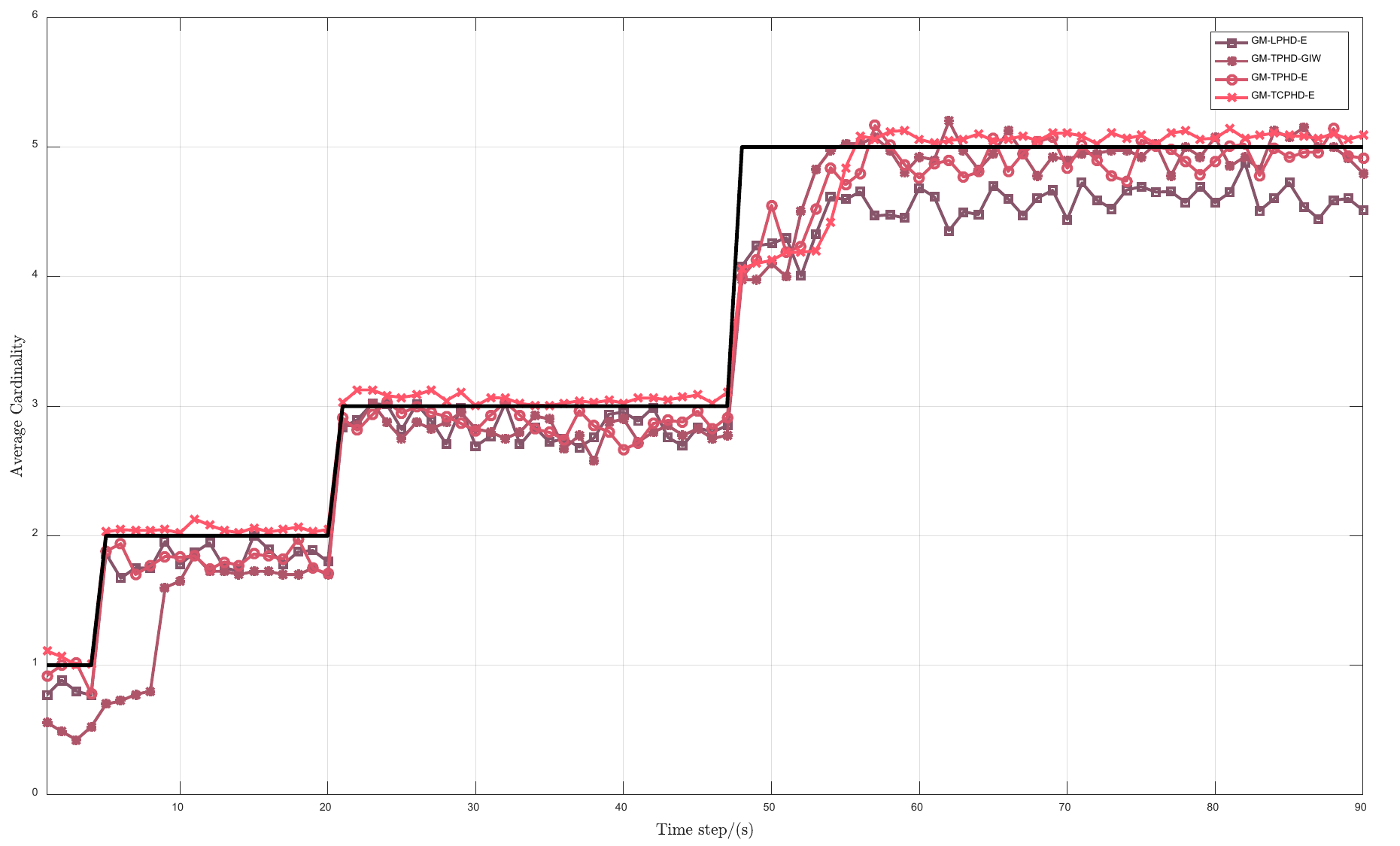}
    \caption{The estimated cardinality result for Scenario 2}
    \label{fig.f2-5}
\end{figure}

We present a sample simulation in Fig.\ref{fig:f2-*} to observe the properties of each method. In addition, to further demonstrate the advantage of the GM-TCPHD-E filter in estimating the number of targets, we introduced cardinality estimation as a metric. The average GWD results, the average TM results, and the average cardinality estimations obtained from 100 MCs using each method are shown in Fig.\ref{fig.f2-3}, Fig.\ref{fig.f2-4}, and Fig.\ref{fig.f2-5}, respectively. Similarly, we recorded the time consumption and the overall TM error under different parameters. The results are presented in Tab.\ref{Tab.f1-1} and Tab.\ref{Tab.f1-2}, respectively.

The results of Scenario 2 validate not only the applicability of our methods in an actual scene, but also further verify our conclusions in Section \ref{Sec6-D}, including that:
\begin{itemize}
    \item The GM-TPHD-GIW filter has shortcomings in shape estimation, so in Fig.\ref{fig.f2-3}, although its GWD does not diverge, it stabilizes at a larger value. In Fig.\ref{fig.f2-4}, the GM-TPHD-GIW filter has the largest location cost, and due to the lag in trajectory building, peaks appear on its missed cost between 1s and 10s, as well as between 48s and 55s.
    \item The GM-LPHD-E filter has a divergent GWD in Fig.\ref{fig.f2-3}, due to trajectory switching between Tar. 4 and Tar. 5, this phenomenon is consistent with its switch cost in Fig.\ref{fig.f2-4}. And as a result of the trajectory switching, there is a clear peak between 58s and 75s.
    \item The two methods we proposed, the GM-TPHD-E filter and the GM-TCPHD-E filter, also showed similar performance. But based on the cardinality estimation results in Fig.\ref{fig.f2-5}, the GM-TCPHD-E filter estimates the cardinality more accurately than the GM-TPHD-E filter, especially at 5s, 21s and 48s when the number of targets changes.
    \item In terms of time consumption and overall TM error under different simulation parameters, the results are consistent with Scenario 1, they once again demonstrate the advantages of the proposed methods in terms of overall accuracy and computation efficiency.
\end{itemize}

\begin{table*}[!t]
 \caption{THE OVERALL TRAJECTORY METRIC ERROR WITH DIFFERENT SIMULATION PARAMETERS}
    \label{Tab.f1-2}
\begin{tabular}{ccccccccc}
\hline
\multicolumn{1}{|c|}{}          & \multicolumn{4}{c|}{Scenario 1}                                                                                                         & \multicolumn{4}{c|}{Scenario 2}                                                                                                         \\ \hline
\multicolumn{1}{|c|}{}          & \multicolumn{1}{c|}{GM-LPHD-E} & \multicolumn{1}{c|}{GM-TPHD-GIW} & \multicolumn{1}{c|}{GM-TPHD-E} & \multicolumn{1}{c|}{GM-TCPHD-E}    & \multicolumn{1}{c|}{GM-LPHD-E} & \multicolumn{1}{c|}{GM-TPHD-GIW} & \multicolumn{1}{|c|}{GM-TPHD-E} & \multicolumn{1}{c|}{GM-TCPHD-E}    \\ \hline
\multicolumn{1}{|c|}{No change} & \multicolumn{1}{c|}{7.16}      & \multicolumn{1}{c|}{6.34}        & \multicolumn{1}{c|}{5.32}      & \multicolumn{1}{c|}{\textbf{4.73}} & \multicolumn{1}{|c|}{7.38}      & \multicolumn{1}{c|}{6.51}        & \multicolumn{1}{c|}{5.71}      & \multicolumn{1}{c|}{\textbf{4.96}} \\ \cline{2-9} 
\multicolumn{1}{|c|}{$\mathsf{Q}^{\boldsymbol{e}}=\frac{1}{4}\dot{\mathsf{Q}}^{\boldsymbol{e}}$}        & \multicolumn{1}{c|}{6.51}      & \multicolumn{1}{c|}{5.31}        & \multicolumn{1}{c|}{4.29}      & \multicolumn{1}{c|}{\textbf{3.61}} & \multicolumn{1}{c|}{6.59}      & \multicolumn{1}{c|}{5.37}        & \multicolumn{1}{c|}{4.49}      & \multicolumn{1}{c|}{\textbf{3.84}} \\ \cline{2-9} 
\multicolumn{1}{|c|}{$\mathsf{Q}^{\boldsymbol{e}}=4\dot{\mathsf{Q}}^{\boldsymbol{e}}$}        & \multicolumn{1}{c|}{8.78}      & \multicolumn{1}{c|}{8.51}        & \multicolumn{1}{c|}{8.11}      & \multicolumn{1}{c|}{\textbf{7.26}} & \multicolumn{1}{c|}{8.91}      & \multicolumn{1}{c|}{8.79}        & \multicolumn{1}{c|}{8.26}      & \multicolumn{1}{c|}{\textbf{7.56}} \\ \cline{2-9} 
\multicolumn{1}{|c|}{$\mathrm{\lambda}^{\mathrm{C}}=\frac{1}{4}\dot{\mathrm{\lambda}}^{\mathrm{C}}$}        & \multicolumn{1}{c|}{7.14}      & \multicolumn{1}{c|}{5.63}        & \multicolumn{1}{c|}{5.49}      & \multicolumn{1}{c|}{\textbf{4.76}} & \multicolumn{1}{c|}{7.24}      & \multicolumn{1}{c|}{5.78}        & \multicolumn{1}{c|}{5.62}      & \multicolumn{1}{c|}{\textbf{4.83}} \\ \cline{2-9} 
\multicolumn{1}{|c|}{$\mathrm{\lambda}^{\mathrm{C}}=4\dot{\mathrm{\lambda}}^{\mathrm{C}}$}        & \multicolumn{1}{c|}{7.41}      & \multicolumn{1}{c|}{8.32}        & \multicolumn{1}{c|}{5.87}      & \multicolumn{1}{c|}{\textbf{5.01}} & \multicolumn{1}{c|}{7.61}      & \multicolumn{1}{c|}{8.59}        & \multicolumn{1}{c|}{5.94}      & \multicolumn{1}{c|}{\textbf{5.12}} \\ \cline{2-9} 
\multicolumn{1}{|c|}{$\mathsf{p}^{\mathsf{D}}=1$}        & \multicolumn{1}{c|}{7.16}      & \multicolumn{1}{c|}{5.32}        & \multicolumn{1}{c|}{4.81}      & \multicolumn{1}{c|}{\textbf{4.29}} & \multicolumn{1}{c|}{7.21}      & \multicolumn{1}{c|}{5.41}        & \multicolumn{1}{c|}{4.97}      & \multicolumn{1}{c|}{\textbf{4.37}} \\ \cline{2-9} 
\multicolumn{1}{|c|}{$\mathsf{p}^{\mathsf{D}}=0.75$}        & \multicolumn{1}{c|}{7.34}      & \multicolumn{1}{c|}{6.42}        & \multicolumn{1}{c|}{6.11}      & \multicolumn{1}{c|}{\textbf{5.91}} & \multicolumn{1}{c|}{7.46}      & \multicolumn{1}{c|}{6.78}        & \multicolumn{1}{c|}{6.49}      & \multicolumn{1}{c|}{\textbf{6.12}} \\ \cline{2-9} 
\multicolumn{1}{|c|}{$\mathsf{p}^{\mathsf{S}}=0.95$}        & \multicolumn{1}{c|}{7.22}      & \multicolumn{1}{c|}{6.54}        & \multicolumn{1}{c|}{5.34}      & \multicolumn{1}{c|}{\textbf{4.82}} & \multicolumn{1}{c|}{7.42}      & \multicolumn{1}{c|}{6.63}        & \multicolumn{1}{c|}{5.47}      & \multicolumn{1}{c|}{\textbf{5.01}} \\ \cline{2-9} 
\multicolumn{1}{|c|}{$w_{k}^{\beta ,\left( j \right)}=0.05$}        & \multicolumn{1}{c|}{7.19}      & \multicolumn{1}{c|}{6.36}        & \multicolumn{1}{c|}{5.35}      & \multicolumn{1}{c|}{\textbf{4.75}} & \multicolumn{1}{c|}{7.41}      & \multicolumn{1}{c|}{6.54}        & \multicolumn{1}{c|}{5.74}      & \multicolumn{1}{c|}{\textbf{4.98}} \\ \hline
\multicolumn{9}{l}{\textbf{Note}: The result in Tab.\ref{Tab.f1-2} is average over all MCs and all time steps, and the symbol `$\dot{a}$' represents the original value of the parameter $a$}                                                                                                                                                                         \\ \hline
\end{tabular}
\end{table*}

According to all the metrics and all scenarios, the GM-TCPHD-E filter is the best performing filter, followed by the GM-TPHD-E filter. Both filters are superior to the RMM-based GM-TPHD-GIW filters in shape estimation, and can provide more accurate trajectory estimation compared to the labeling-based GM-LPHD-E filter. Overall, performance impairs for all filters if the detection / survival probability decreases, the intensity of birth decreases, the intensity of clutter increases, and the covariance of the measurement noise increases; but our proposed GM-TCPHD-E filter is least affected and exhibits the best robustness.

\section{Conclusion}
\label{Sec7}

In this article, we have developed the PHD filter and the CPHD filter and extended them to METT applications that can accurately estimate the shape of targets, and combined them with TST to facilitate the complete and stable establishment of target trajectories. We further provide the Gaussian mixture implementations with closed-form recursions for the proposed TPHD-E and TCPHD-E filters. Moreover, the proposed methods have been comprehensively examined using multiple metrics, including the Gaussian-Wasserstein distance, the trajectory metric, and the computational cost. Overall, the proposed GM-TCPHD-E filter has the best performance and both proposed methods provide better performances in target shape estimation, target trajectory building, and computational consumption compared to existing methods.

Although the PHD filter and the CPHD filter have significant advantages in computational consumption and scalability, their applicability is limited due to their being driven by the target birth model rather than driven by measurement. Therefore, combining our research with more advanced RFS-based filters, such as the extended target Poisson multi-Bernoulli mixture filter\cite{2019GranströmTAES,2022XiaTAES}, will become our future work.

\bibliographystyle{IEEEtran}
 
\bibliography{ref.bib}





\end{document}